\documentclass[aps]{revtex4}
\usepackage{amsfonts}
\usepackage{amsmath}
\usepackage{amssymb,epsf}

\begin{document}

\title{Phase transition of charged black holes in massive gravity through
new methods}
\author{S. H. Hendi$^{1,2}$\footnote{%
email address: hendi@shirazu.ac.ir}, S. Panahiyan$^{1,3}$\footnote{%
email address: sh.panahiyan@gmail.com}, B. Eslam Panah$^{1}$\footnote{%
email address: behzad.eslampanah@gmail.com} and M. Momennia$^{1}$\footnote{%
email address: momennia1988@gmail.com}} \affiliation{$^1$ Physics
Department and Biruni Observatory, College of Sciences, Shiraz
University, Shiraz 71454, Iran\\
$^2$ Research Institute for Astronomy and Astrophysics of Maragha
(RIAAM), P.O. Box 55134-441, Maragha, Iran\\
$^3$ Physics Department, Shahid Beheshti University, Tehran 19839,
Iran}

\begin{abstract}
Motivated by providing preliminary steps to understand the
conception of quantum gravity, in this paper, we study the phase
structure of a semiclassical gravitational system. We investigate
the stability conditions and phase transition of charged black
holes in massive gravity via canonical and geometrical
thermodynamic approaches. We point out the effects of massive
parameter on stability conditions of these black holes and show
how massive coefficients affect the phase transition points of
these black holes. We also study the effects of boundary topology
on thermodynamical behavior of the system. In addition, we give
some arguments regarding the role of higher dimensions and
highlight the effect of the electric charge in thermodynamical
behavior. Then, we extend our study to geometrical thermodynamic
approach and show that it can be a successful method for studying
the black hole phase transition. At last, by employing the
relation between thermodynamical pressure and cosmological
constant, critical behavior of the system and the effects of
different parameters on critical values are investigated.
\end{abstract}

\maketitle

\section{Introduction}

Cosmological observations show that, $95\%$ of the Universe is
made of dark energy and dark matter. In order to interpret dark
sector of the Universe, some various candidates have been proposed
in literature, such as modified gravities like Lovelock gravity
\cite{LovelockI,LovelockII,Deruelle}, brane world cosmology
\cite{Brax,Gergely,Demetrian,Bouhmadi,Cruz}, scalar-tensor
theories \cite%
{Brans,CaiM,Sotiriou,DehghaniP,HendiJM,Maeda,HendiK,Gao,Huang,Crisostomi},
and the so-called $F(R)$ gravity theories \cite%
{Akbar,Souza,Atazadeh,Cognola,Corda,SotiriouF,Hendi,NojiriO,HendiEM,HendiMRE,HendiII,Zhang,HendiEC,Miao,HendiES,CapGLP,CapGV}
(for a review regarding thermodynamical aspects of modified
theories, see Ref. \cite{REV}). One of the interesting aspects of
these
theories is graviton. Graviton in the Einstein gravity is massless \cite%
{massless1,massless2,massless3,massless4}. A natural question is whether one
can build a self-consistent gravity theory if graviton is massive \cite%
{dvali1,dvali2,dvali3}. To do so, one should modify the Einstein
theory of gravity to include massive gravitons \cite
{dvali1,dvali2,dvali3,NEW,bi,dRGT1,dRGT2,dRGT3}. From the
theoretical perspective, the shear difficulty of constructing a
consistent theory of massive gravity makes the problem more
interesting.

Fierz and Pauli developed the ghost-free theory of non-interacting massive
gravitons which had a condition of being linearized over Minkowski space
\cite{Fierz,FierzP}. Then, Boulware and Deser showed that a generic
extension of the Fierz-Pauli theory to curved backgrounds will contain ghost
instabilities \cite{Boulware}. In Refs. \cite%
{dRGT1,dRGT3,Rham,HassanI,Hassan,HassanII}, a potentially
ghost-free non-linear massive gravity actions has been
constructed. Furthermore, one of the classes of charged black hole
solutions in nonlinear massive gravity was studied in Ref.
\cite{Saridakis}. In addition, the effective field theory of the
modified gravity with massive modes has been studied in Ref.
\cite{effect}. The phenomenological aspects of massive gravities
with Lorentz-violation and invariant mass terms were investigated
\cite{infr}. Furthermore, massive, massless and ghost modes of
gravitational waves in higher order gravities were studied and the
possibility of the detection and additional polarization modes of
a stochastic gravitational wave were discussed in Ref.
\cite{waves}. In a series of studies, the black
hole solutions in different massive gravities \cite%
{massiveblack1,massiveblack2,massiveblack3,massiveblack4,massiveblack5,massiveblack6,massiveblack7,massiveblack8,massiveblack9,massiveblack10,massiveblack11}
have been investigated and it was shown that some of these black
holes may enjoy anti-evaporation in their thermodynamical
properties \cite{anti1,anti2}. The massive gravity has also been
employed to conduct some studies in the context of cosmology (for
an incomplete list of these studies, we refer the reader to Refs.
\cite
{cosmology1,cosmology2,cosmology3,cosmology4,cosmology5,cosmology6,cosmology7,cosmology8,cosmology9}).
Of the other applications of massive gravity, one can point out
the large applications of this gravity in the anti-de
Sitter/conformal field theory (AdS/CFT) community. For example,
the holographical aspects of the massive gravity coupled with
nonlinear electrodynamic models have been investigated in Ref.
\cite{holo1}. Furthermore, the extension of massive gravities with
additional couplings and higher derivative terms have been
addressed and their holographical properties are extracted
\cite{holo2,holo3,holo4,holo5}.

Recently, David Vegh \cite{Vegh} found a nontrivial black hole
solution with a Ricci flat horizon in four dimensional massive
gravity with a negative cosmological constant. He studied an
assumption that the graviton may have lattice like behavior in the
holographic conductor model: the conductivity generally exhibits a
Drude peak which approaches to a delta function in the massless
gravity limit. Application of massive gravity in the context of
AdS/CFT correspondence via stability conditions
and metal-insulator transition has been studied in \cite%
{MassiveADSCFT,Taylor,BaggioliI,BaggioliII}. In addition,
thermodynamical behavior and $P-V$ criticality of the Vegh's
massive gravity have been, recently, investigated
\cite{Cai2015,Xu2015,HendiEPJI,HendiPEJII,HendiEPJIII}. Some
holographic effects of graviton mass have been investigated in
Ref. \cite{Davison,Blake}. The main goal of this paper is to
obtain Vegh's black hole solutions and study corresponding
thermodynamical properties and their phase structure.

Thermodynamical aspects of the black holes and their critical
behavior have been one of the greatest interests for many authors.
Recently, it was proposed to change the role of some constants to
the cases of (dynamical or
thermodynamical) variables \cite%
{Kastor,DolanI,Gibbons,Rasheed,Breton,Huan,HendiV,DavisonSZ,HendiPEPTEP}.
Among the thermodynamical properties of the black holes, thermal
stability and phase transition have been investigated in
literature. Studying thermal stability and phase transition points
can be done in the context of canonical ensemble by analyzing the
heat capacity. The sign of heat capacity represents
stability/instability of the black holes, while its divergencies
and roots are denoted as second order phase transition and bound
points, respectively.

Another approach toward studying critical behavior of the black
holes is through geometrical method. In this case, by employing a
thermodynamical potential and corresponding extensive parameters,
one can construct a phase space containing information regarding
phase transition of the black holes. In other words, geometrical
thermodynamics tries to introduce thermodynamical properties of
the systems through the use of the Riemannian geometry. The basic
motivation is to give an independent picture regarding
thermodynamical aspects of systems. In addition, geometrical
thermodynamics gives information regarding bound points, phase
transitions, their types and stability conditions. Furthermore, it
contains information regarding molecular interaction around phase
transitions for the mundane systems. In other words, by studying
the sign of thermodynamical Ricci scalar (TRS) around phase
transition points, one can extract information whether interaction
is repulsive or attractive. There are several approaches in the
context of geometrical thermodynamics (GTs); the well-known
metrics of the Weinhold \cite{Weinhold,WeinholdII}, Ruppeiner
\cite{Ruppeiner,RuppeinerII} and Quevedo metric which were
proposed in different structures. Recently it was pointed out that
there may be some mismatches between second order phase
transitions and divergencies of TRS in the mentioned methods
\cite{Shen,Mirza,Aman,Sarkar}, and an alternative thermodynamical
metric was introduced to solve this problem \cite{New Metric,New
MetricII,New MetricIII,New MetricIV}. The generalization of this
metric in the context of extended phase space (in which constants
considered as variables) was done in Ref. \cite{HPE}. It was shown
that thermodynamical behavior of the system in the context of
extended phase space, heat capacity and GTs lead to same
consequences \cite{HPE}. The importance of geometrical methods
comes from the fact that these methods are based on the grand
canonical ensemble foundation. Although we connect this method to
the heat capacity, it is worthwhile to mention that this is only
for the sake of comparison and check the validity of the results.

Recent progresses in the AdS/CFT correspondence, suggest that cosmological
constant should be considered as a thermodynamical variable \cite%
{CosVar2,CosVar3,CosVar4,CosVar5}. In addition, in black hole
thermodynamics, it was shown that such consideration could lead to remove
specific thermodynamical problems such as ensemble dependency \cite{Mamasani}%
. One of the corresponding thermodynamical variable for
cosmological constant is pressure. It was shown that by using this
analogy, one can introduce novel properties in black hole
thermodynamics such as Van der
Waals like behavior, triple point and reentrant of phase transition \cite%
{CosmP2,CosmP4,CosmP5,CosmP6,CosmP7,CosmP8,CosmP10,CosmP11,CosmP12,CosmP13}.
Therefore, by such consideration, it is possible to study critical behavior
of the black holes in extended phase space. One of the new methods for
extracting critical points of the black holes was proposed and employed in
Ref. \cite{HPE}. In this method, by using proportionality between
cosmological constant and pressure in denominator of the heat capacity, a
relation is obtainable for pressure which is independent of equation of
state. In this paper, we will employ this method to extract critical points
of the black hole system.

The outline of the paper is as follow; in section II, we review
the charged massive black holes and their thermodynamical
quantities. In section III, we introduce approaches for studying
phase transition of these black holes in the context of heat
capacity and geometrical thermodynamics, and study thermal
stability of these black holes. Then, we investigate the existence
of the phase transitions in the context of two mentioned
approaches. The last section is devoted to closing remarks.

\section{Black hole solutions in massive gravity}

The ($n+2$)-dimensional action of massive gravity with negative cosmological
constant can be written as \cite{Cai2015,Xu2015,HendiEPJI,HendiPEJII}
\begin{equation}
\mathcal{I}=\frac{1}{2\mathcal{\kappa }^{2}}\int d^{n+2}x\sqrt{-g}\left[
\mathcal{R}+\frac{n(n+1)}{l^{2}}-\frac{\mathcal{F}}{4}+m^{2}%
\sum_{i=1}^{n+2}c_{i}\mathcal{U}_{i}(g,f)\right] ,  \label{Action}
\end{equation}%
where $\mathcal{R}$ is the scalar curvature, $\mathcal{F}$ is the
Maxwell invariant, $f$ is a fixed rank-$2$ symmetric tensor and
$m^{2}$ is the positive massive parameter (see Ref.
\cite{Vegh,Davison1} for more details regarding the relation of
positive $m^{2}$ with the wall of
stability interpretation). In addition, $c_{i}$'s are constants and $%
\mathcal{U}_{i}$'s are symmetric polynomials of the eigenvalues of the $%
(n+2)\times (n+2)$ matrix $\mathcal{K}_{\nu }^{\mu }=\sqrt{g^{\mu
\alpha }f_{\alpha \nu }}$ \cite{higher1}
\begin{eqnarray}
\mathcal{U}_{1} &=&\left[ \mathcal{K}\right] ,  \nonumber \\
\mathcal{U}_{2} &=&\left[ \mathcal{K}\right] ^{2}-\left[ \mathcal{K}^{2}%
\right] ,  \nonumber \\
\mathcal{U}_{3} &=&\left[ \mathcal{K}\right] ^{3}-3\left[ \mathcal{K}\right] %
\left[ \mathcal{K}^{2}\right] +2\left[ \mathcal{K}^{3}\right] ,  \nonumber \\
\mathcal{U}_{4} &=&\left[ \mathcal{K}\right] ^{4}-6\left[ \mathcal{K}^{2}%
\right] \left[ \mathcal{K}\right] ^{2}+8\left[ \mathcal{K}^{3}\right] \left[
\mathcal{K}\right] +3\left[ \mathcal{K}^{2}\right] ^{2}-6\left[ \mathcal{K}%
^{4}\right] ,  \nonumber \\
\mathcal{U}_{5} &=&\left[ \mathcal{K}\right] ^{5}-10\left[ \mathcal{K}\right]
^{3}\left[ \mathcal{K}^{2}\right] +20\left[ \mathcal{K}\right] ^{2}\left[
\mathcal{K}^{3}\right] -20\left[ \mathcal{K}^{2}\right] \left[ \mathcal{K}%
^{3}\right] +15\left[ \mathcal{K}\right] \left[ \mathcal{K}^{2}\right]
^{2}-30\left[ \mathcal{K}\right] \left[ \mathcal{K}^{4}\right] +24\left[
\mathcal{K}^{5}\right] ,  \nonumber \\
&&...\;\; .
\end{eqnarray}

The square root in $\mathcal{K}$ means $\left( \sqrt{A}\right) _{\nu }^{\mu
}\left( \sqrt{A}\right) _{\lambda }^{\nu }=A_{\lambda }^{\mu }$ and $\left[
\mathcal{K}\right] =\mathcal{K}_{\mu }^{\mu }$. Variation of the action (\ref%
{Action}) with respect to the metric tensor $g_{\mu \nu }$ and the Faraday
tensor $F_{\mu \nu }$, leads to
\begin{equation}
G_{\mu \nu }-\frac{n(n+1)}{2l^{2}}g_{\mu \nu }=\frac{1}{2}\left[ F_{\mu
\lambda }F_{\nu }^{\lambda }-\frac{1}{4}g_{\mu \nu }\mathcal{F}\right]
+m^{2}\chi _{\mu \nu },  \label{Field equation}
\end{equation}%
\begin{equation}
\nabla _{\mu }F^{\mu \nu }=0,  \label{Maxwell equation}
\end{equation}%
where $G_{\mu \nu }$ is the Einstein tensor and
\begin{eqnarray}
\chi _{\mu \nu } &=&-\frac{c_{1}}{2}\left( \mathcal{U}_{1}g_{\mu \nu }-%
\mathcal{K}_{\mu \nu }\right) -\frac{c_{2}}{2}\left( \mathcal{U}_{2}g_{\mu
\nu }-2\mathcal{U}_{1}\mathcal{K}_{\mu \nu }+2\mathcal{K}_{\mu \nu
}^{2}\right) -\frac{c_{3}}{2}(\mathcal{U}_{3}g_{\mu \nu }-3\mathcal{U}_{2}%
\mathcal{K}_{\mu \nu }+  \nonumber \\
&&6\mathcal{U}_{1}\mathcal{K}_{\mu \nu }^{2}-6\mathcal{K}_{\mu \nu }^{3})-%
\frac{c_{4}}{2}(\mathcal{U}_{4}g_{\mu \nu }-4\mathcal{U}_{3}\mathcal{K}_{\mu
\nu }+12\mathcal{U}_{2}\mathcal{K}_{\mu \nu }^{2}-24\mathcal{U}_{1}\mathcal{K%
}_{\mu \nu }^{3}+24\mathcal{K}_{\mu \nu }^{4})+... \; .
\end{eqnarray}

We take into account the metric of ($n+2$)-dimensional spacetime with the
following form
\begin{equation}
ds^{2}=-f(r)dt^{2}+f^{-1}(r)dr^{2}+r^{2}h_{ij}dx_{i}dx_{j},\ i,j=1,2,3,...,n
\end{equation}%
where $h_{ij}dx_{i}dx_{j}$ is the line element for an
$n-$dimensional Einstein space with constant curvature $n(n+1)k$
and volume $V_{n}$. We should note that the constant $k$ indicates
that the boundary of $t=constant$ and $r=constant$ can be a
positive (elliptic), zero (flat) or negative (hyperbolic) constant
curvature hypersurface.

A generalized version of $f_{\mu \nu }$ was proposed in Ref. \cite%
{Cai2015,Xu2015,HendiEPJI,HendiPEJII} with the following form
\begin{equation}
f_{\mu \nu }=diag(0,0,c_{0}^{2}h_{ij}),
\end{equation}%
where by employing it, the metric function $f(r)$ will be \cite%
{Cai2015,Xu2015}
\begin{eqnarray}
f(r) &=&k+\frac{r^{2}}{l^{2}}-\frac{m_{0}}{r^{n-1}}+\frac{q^{2}}{%
2n(n-1)r^{2(n-1)}}+\frac{c_{0}c_{1}m^{2}}{n}r+c_{0}^{2}c_{2}m^{2}+  \nonumber \\
&&(n-1)c_{0}^{3}c_{3}m^{2}r^{-1}+(n-1)(n-2)c_{0}^{4}c_{4}m^{2}r^{-2}+...\;
,
\end{eqnarray}
in which $m_{0}$ and $q$ are integration constants which are
related to the total mass and electric charge of the black hole,
respectively. As one can see, the massive terms could be up to
$n+2$ number. Technically, it is not easy to study the total
behavior of the solutions with this number of terms. Therefore,
for the simplicity, we restrict
our study to $\mathcal{U}_{i}$ up to the fourth term, $\mathcal{U}%
_{4}$ \footnote{%
The similar situation could be observed for generalization of the
Einstein gravity to the Lovelock theory. The effective gravity in
$4$ dimensional case is Einstein theory. In $5$ dimensions, the
contributions of the Gauss-Bonnet terms appear and etc. But, it is
a usual practice to study higher dimensional Einstein/Gauss-Bonnet
gravity without consideration of the higher terms of
Lovelock gravity. Here too, we have restricted our study up to $\mathcal{U}%
_{4}$.}.

\begin{figure}[tbp]
$%
\begin{array}{cc}
\epsfxsize=8.5cm \epsffile{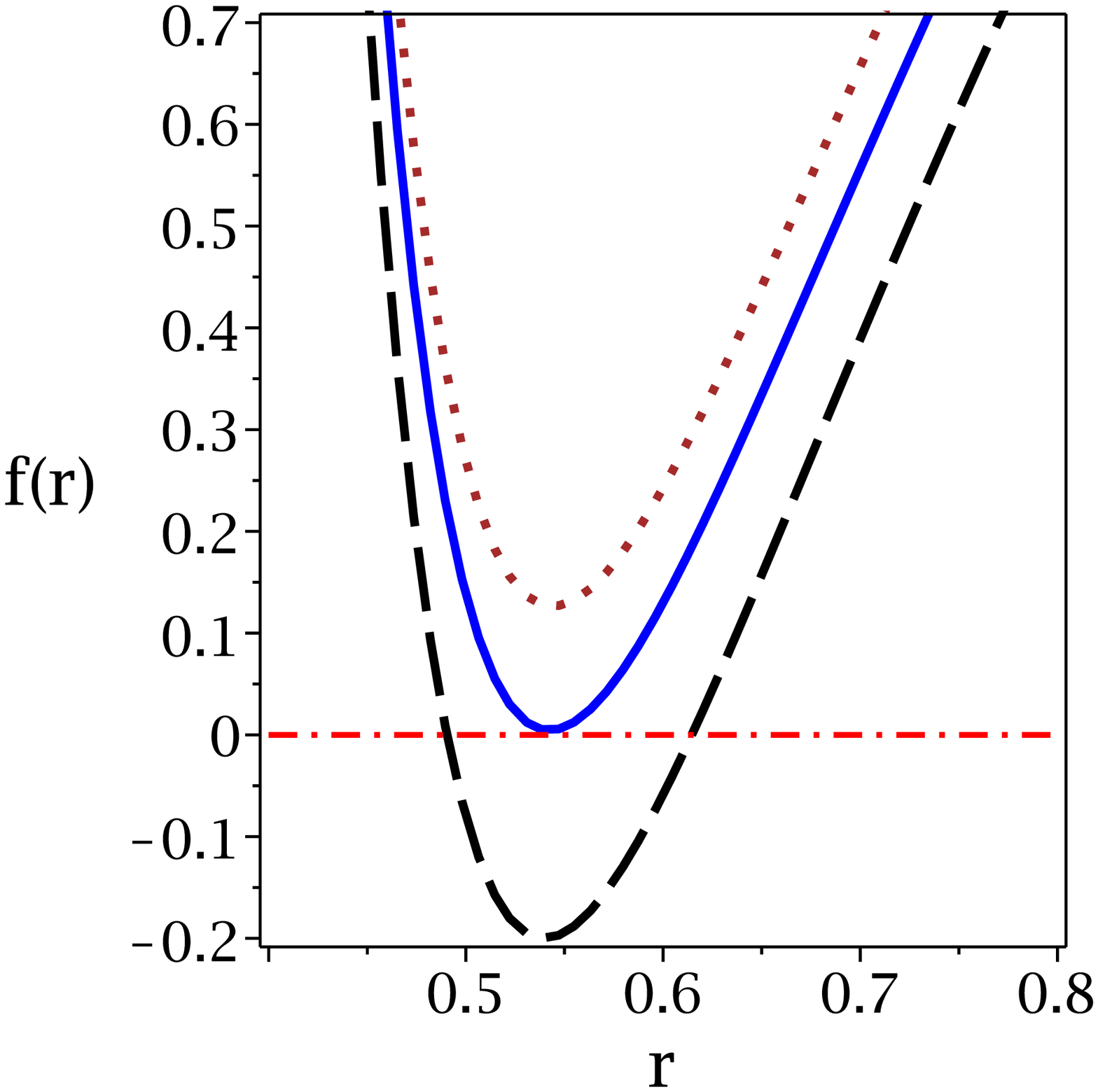} & \epsfxsize=8.5cm %
\epsffile{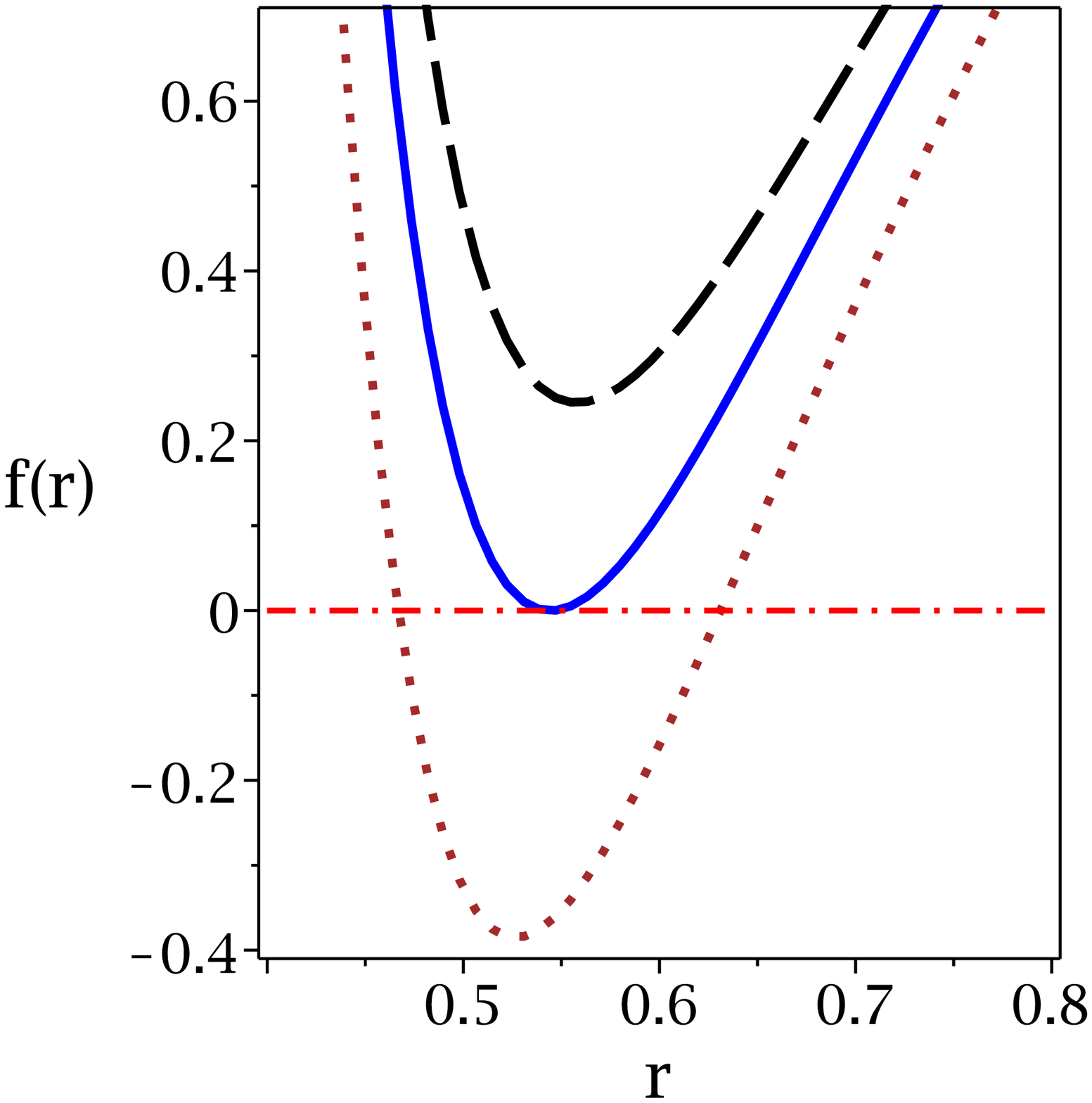}%
\end{array}
$%
\caption{ $f(r)$ versus $r$ for $l=1$, $q=1$, $c_{0}=0.1$, $c_{1}=-0.3$, $%
c_{2}=2$, $c_{3}=3$, $c_{4}=2$, $k=1$ and $n=4$. \newline
Left diagram for $m_{0}=0.5$, $m=0.3$ (dashed line), $m=2.38$ (continues
line) and $m=3$ (dotted line). \newline
Right diagram for $m=2$, $m_{0}=0.45$ (dashed line), $m_{0}=0.49$ (continues
line) and $m_{0}=0.55$ (dotted line).}
\label{Figfrtwo}
\end{figure}


\begin{figure}[tbp]
\epsfxsize=9cm \epsffile{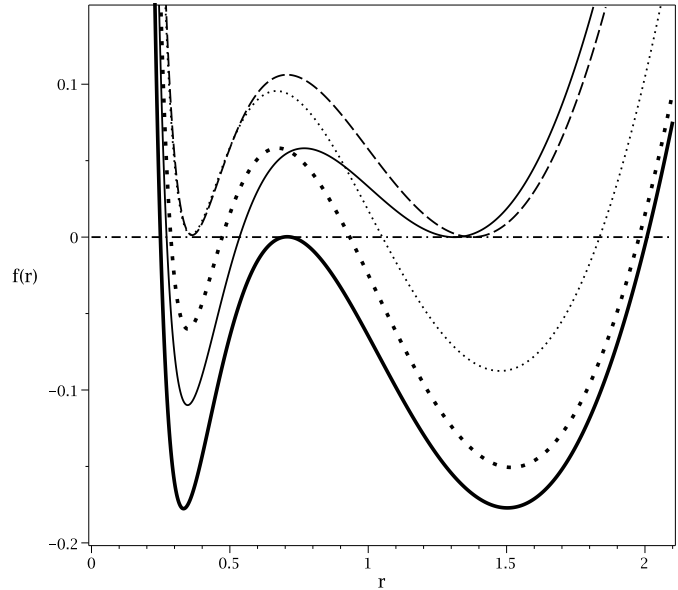}
\caption{$f(r)$ versus $r$ for $l=1$, $q=1$, $c_{0}=0.1$, $c_{1}=-2.3$, $%
c_{2}=10$, $k=1$, and $n=2$. \newline
Diagrams for $m_{0}=1.648$, $m=5.50$ (dashed line), $m_{0}=1.668$, $m=5.60$
(dotted line), $m_{0}=1.70$, $m=5.65$ (bold-dotted line), $m_{0}=1.68$, $%
m=5.462$ (continues line) and $m_{0}=1.74$, $m=5.65$ (bold line). }
\label{Figfr4}
\end{figure}

\begin{figure}[tbp]
$%
\begin{array}{cc}
\epsfxsize=6.5cm \epsffile{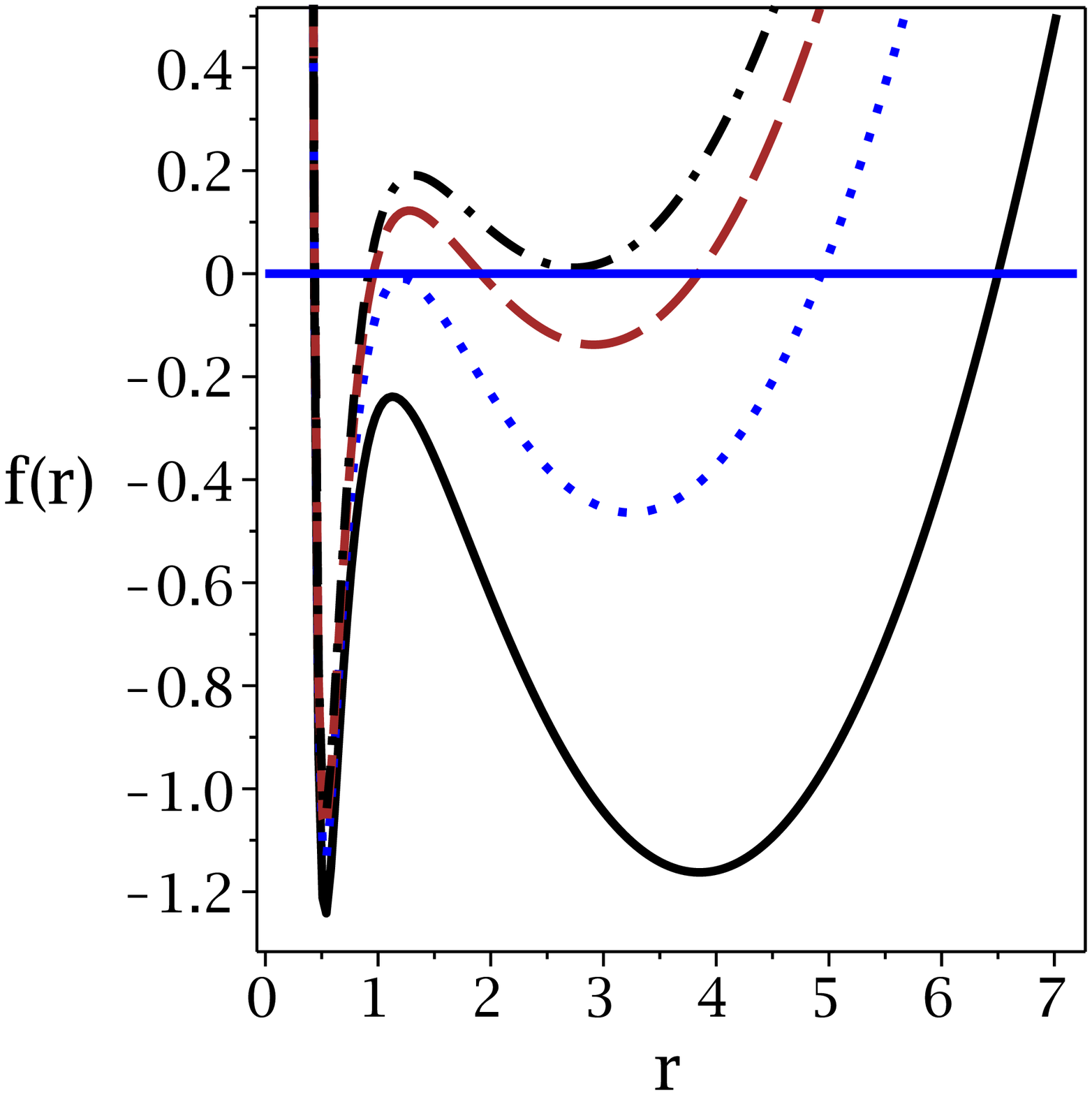} & \epsfxsize=6.5cm %
\epsffile{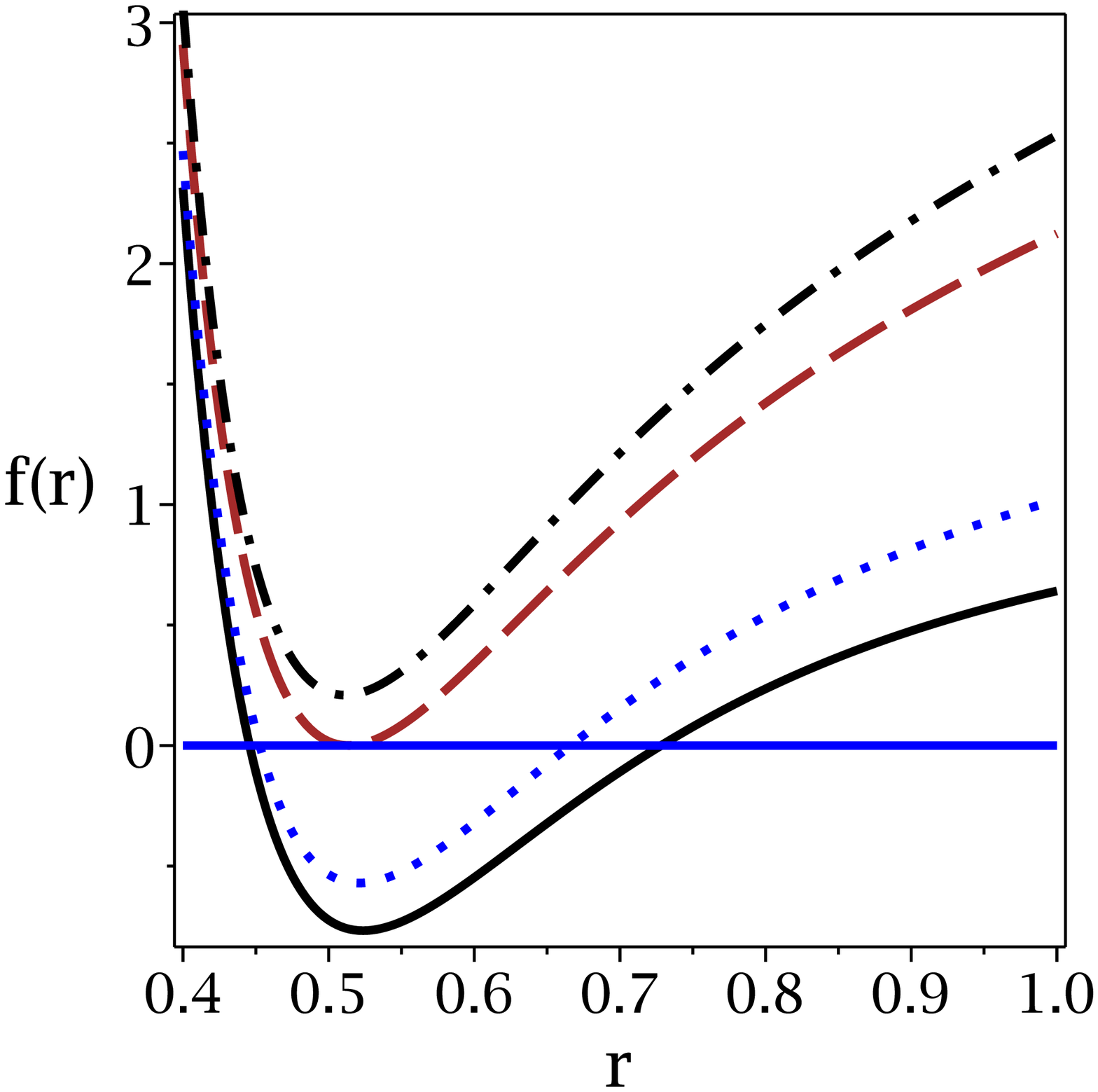} \\
\epsfxsize=6.5cm \epsffile{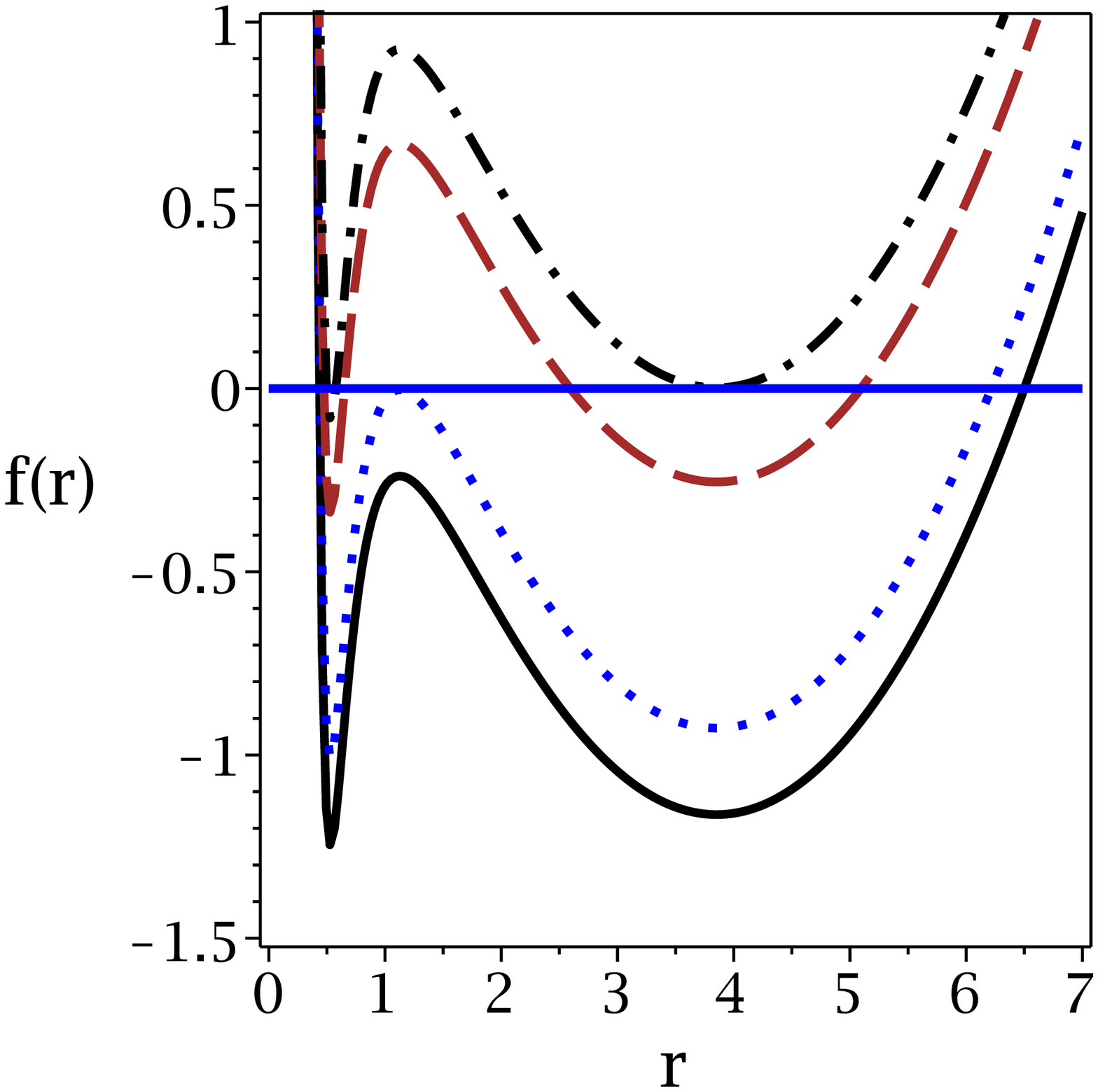} & \epsfxsize=6.5cm %
\epsffile{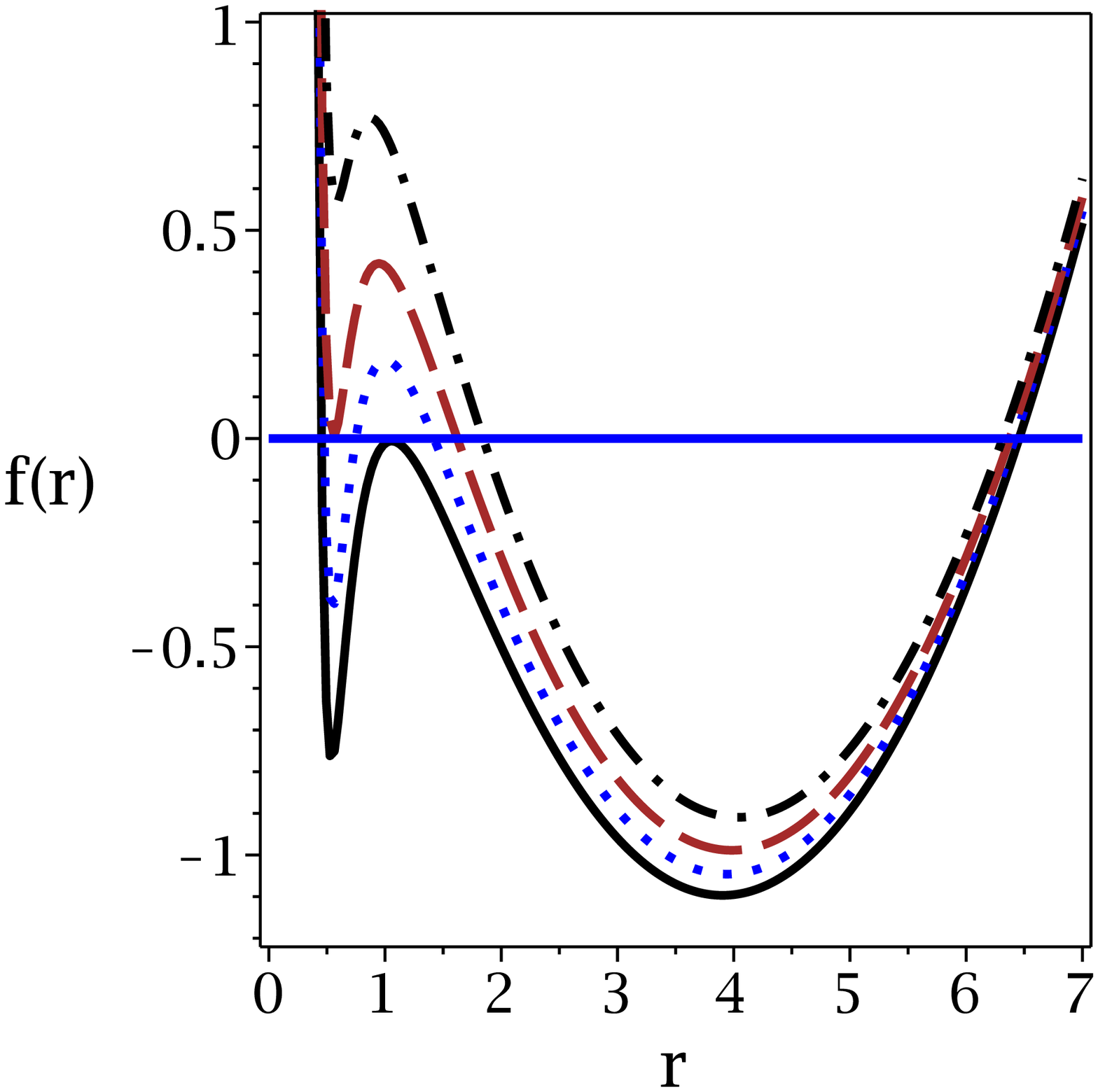}%
\end{array}
$%
\caption{ $f(r)$ versus $r$ for $l=1$, $q=1$, $c_{0}=0.1$,
$c_{4}=1$, $k=1$, $m_{0}=0.6$, $m=5.5$ and $n=4$. \newline
Left-up panel: for $c_{2}=c_{3}=1$, $c_{1}=-1.7$ (continuous line), $%
c_{1}=-1.44$ (dotted line), $c_{1}=-1.3$ (dashed line) and
$c_{1}=-1.23$ (dashed-dotted line). \newline Right-up panel: for
$c_{2}=c_{3}=1$, $c_{1}=-0.5$ (continuous line), $c_{1}=0$ (dotted
line), $c_{1}=1.46$ (dashed line) and $c_{1}=2$ (dashed-dotted
line). \newline
left-down panel: for $c_{1}=-1.7$, $c_{3}=1$, $c_{2}=1$ (continuous line), $%
c_{2}=1.78$ (dotted line), $c_{2}=4$ (dashed line) and $c_{2}=4.85$
(dashed-dotted line). \newline
right-down panel: for $c_{1}=-1.7$, $c_{2}=1$, $c_{3}=3.8$ (continuous line), $%
c_{3}=6$ (dotted line), $c_{3}=8.5$ (dashed line) and $c_{3}=12$
(dashed-dotted line).} \label{Figfr44}
\end{figure}


In order to study the effects of massive gravity, one can investigate the
metric function. Regarding various terms of $f(r)$, it is worthwhile to
mention that fourth term ($q^2$ term) is dominant near the origin and
therefore one can conclude that the singularity is timelike. In addition,
for large distance, second term ($r^2/l^2$ term) is dominant which confirms
that the solutions are asymptotically adS. It is evident that the behavior
of the metric function is highly sensitive to massive parameters and
contribution of the massive gravity (see Figs. \ref{Figfr4} and \ref{Figfr44}%
). For specific values of different parameters, the total behavior of the
metric function will be similar to the metric function of Einstein-Maxwell
gravity which is containing two roots, one extreme root or without any root
(see Fig. \ref{Figfrtwo}). Whereas by considering other choices of free
parameters, the following configurations for the roots of the metric
function may happen. For example, two extreme roots (dashed line in Fig. \ref%
{Figfr4}), three roots in which the smaller one is extreme (dotted line in
Fig. \ref{Figfr4}), four roots (bold-dotted line in Fig. \ref{Figfr4}),
three roots where the bigger one is extreme (continues line in Fig. \ref%
{Figfr4}) and three roots which the middle one is extreme (bold
continues line in Fig. \ref{Figfr4}). As one can see, these
different behaviors are originated from contributions of the
massive gravity. The existence
of multi-horizon for massive gravity has been reported in Refs. \cite%
{anti1,anti2} as well. It was shown that the existence of such
property with specific quantum corrections leads to a phenomena
which is opposite of the black holes evaporation. In other words,
there might exist an anti-evaporation behavior for the black holes
in these cases. Here, we see that similar property is observed for
these black holes. Therefore, it is possible that anti-evaporation
property exists for these black holes in the presence of massive
gravity. It is worthwhile to mention that anti-evaporation has
been reported for specific black holes in $F(R)$ gravity as well
\cite{anti3,anti4,anti5}.

Considering the fact that we are interested in studying GTs and heat
capacity of the solutions, we give a brief review about the conserved
charges and thermodynamic quantities in which they have been obtained in
Ref. \cite{Cai2015}.

Regarding the obtained solutions, the action (\ref{Action})
diverges, as is the Hamiltonian and other associated conserved
quantities. One of the systematic methods for removing this
divergency is through the use of the counterterm method inspired
by AdS/CFT correspondence \cite{Mal1,Mal2,Mal3,Mal4}. It was also
shown that by using the Hamiltonian approach, one can find the
mass $M$ of the black hole solutions as ( see Ref. \cite{Cai2015}
for more details)
\begin{equation}
M=\frac{nV_{n}}{2\mathcal{\kappa }^{2}}m_{0}.  \label{Mass}
\end{equation}%
In addition, the electric charge of the black hole, $Q$, can be
obtained by using the Gauss's law and calculating the
electromagnetic
flux at infinity. Straightforward calculations lead to the following result%
\begin{equation}
Q=\frac{V_{n}}{2\mathcal{\kappa }^{2}}q.  \label{Charge}
\end{equation}

The temperature may be obtained through the use of regularity of
the solutions at the event horizon (the largest root of metric
function which is denoted by $r_{+}$). It is also calculated
through the use of the surface gravity definition as

\begin{equation*}
T=\frac{1}{2\pi }\sqrt{-\frac{1}{2}\left( \nabla _{\mu }\chi _{\nu }\right)
\left( \nabla ^{\mu }\chi ^{\nu }\right) },
\end{equation*}%
where $\chi =\partial _{t}$ is the Killing vector. It is easy to
show that \cite{Cai2015}
\begin{eqnarray}
T &=&\frac{1}{4\pi r_{+}}\left[ (n-1)k+(n+1)\frac{r_{+}^{2}}{l^{2}}-\frac{%
q^{2}}{2nr_{+}^{2(n-1)}}+c_{1}c_{0}m^{2}r_{+}(n-1)c_{2}c_{0}^{2}m^{2}+%
\right.   \nonumber \\
&&\left. \frac{(n-1)(n-2)c_{3}c_{0}^{3}m^{2}}{r_{+}}+\frac{%
(n-1)(n-2)(n-3)c_{4}c_{0}^{4}m^{2}}{r_{+}^{2}}\right] ,  \label{temp}
\end{eqnarray}%
where $r_{+}$ satisfy $f(r=r_{+})=0$.

At last, we should obtain the entropy of the black hole solutions.
The entropy of the black holes satisfies the so-called area law \cite%
{Beck1,Beck2,Beck3,Beck4,Beck5,Beck6} in the context of Einstein
gravity. According to this law, one finds that the black hole
entropy is equal to one-quarter of the horizon area, i.e.,
\begin{equation}
S=\frac{2\pi V_{n}}{\mathcal{\kappa }^{2}}r_{+}^{n}.  \label{Entropy}
\end{equation}

\section{Geometrical study of the phase transition and thermal stability}

In this section, we study the local stability and second order
phase transition of the solutions in the canonical ensemble and
compare our results with a new GTs approach for the phase
transition and bound points.

In the canonical ensemble, the positivity of the heat capacity is sufficient
to ensure thermal stability. One can calculate the heat capacity as
\begin{equation}
C_{Q}=T\left( \frac{\partial S}{\partial T}\right) _{Q}=\frac{\left( \frac{%
\partial M}{\partial S}\right) _{Q}}{\left( \frac{\partial ^{2}M}{\partial
S^{2}}\right) _{Q}}.  \label{heatcap}
\end{equation}

Considering the fact that changing in the sign of heat capacity is
representing the phase transition between unstable/stable states,
we regard divergence points of the heat capacity as second order
phase transition points. So, the second order phase transition and
bound points of the black holes with regular $T$ in the context of
heat capacity are indicated with following relations
\begin{equation}
\left\{
\begin{array}{cc}
T=\left( \frac{\partial M}{\partial S}\right) _{Q}=0 & bound\;\;point \\
&  \\
\left( \frac{\partial ^{2}M}{\partial S^{2}}\right) _{Q}=0 & phase
\;\;transition \;\; point
\end{array}%
\right. ,  \label{phase}
\end{equation}%
where we called the roots and divergence points of the heat
capacity as bound points and (second order) phase transition
points, respectively. It is notable that $T=0$ indicates a bound
point between nonphysical ($T<0$) and physical ($T>0$) region.

On the other hand, GTs is another way for investigating the phase
transition in the context of black hole thermodynamics. In this
approach, several metrics have been introduced (Ruppeiner,
Weinhold and Quevedo metrics) in which by using of these metrics,
one can investigate the phase transition of black holes. It was
previously shown that these metrics encounter with some problems
for specific types of black holes \cite{New Metric,New
MetricII,New MetricIII,New MetricIV}. Recently, a new metric (HPEM
metric) was proposed in order to solve the problems that other
metrics may confront \cite{New Metric,New MetricII,New
MetricIII,New MetricIV}. The roots of denominator of the Ricci
scalar of HPEM metric only contains the roots of numerator and
denominator of the heat capacity. In other words, divergence
points of the Ricci scalar of HPEM metric only coincide with the
roots and divergence points of the heat capacity. The HPEM metric
has the following form
\begin{equation}
ds_{HPEM}^{2}=\frac{SM_{S}}{\left( \Pi _{i=2}^{n}\frac{\partial ^{2}M}{%
\partial \chi _{i}^{2}}\right) ^{3}}\left(
-M_{SS}dS^{2}+\sum_{i=2}^{n}\left( \frac{\partial ^{2}M}{\partial \chi
_{i}^{2}}\right) d\chi _{i}^{2}\right) ,  \label{HPEM}
\end{equation}%
where $M_{S}=\partial M/\partial S$, $M_{SS}=\partial
^{2}M/\partial S^{2}$ and $\chi _{i}$ ($\chi _{i}\neq S$) are
extensive parameters. In what follows, we will study the stability
and second order phase transition of the charged massive black
holes in the context of heat capacity and GTs.

First of all, one should take this fact into consideration that the sign of
temperature is putting a restriction on systems to be physical or
non-physical. As we see later, there will be a critical horizon radius, $%
r_{+c}$, in which for $r_{+}<r_{+c}$ the temperature of the system is
negative and in this region solutions are non-physical.

It is a matter of calculation to show that for the root of the heat
capacity, one can find following relation with respect to the massive
parameter
\begin{equation}
m_{c}\equiv \left. m\right\vert _{C_{Q}=0}=\frac{1}{2l}\sqrt{\frac{A}{B}},
\label{mzero}
\end{equation}%
where%
\begin{equation}
A=2\left[ q^{2}l^{2}-2n\left( n+1\right) r_{+}^{2n}\right]
r_{+}-4nl^{2}kr_{+}^{2n-4},
\end{equation}%
\begin{equation}
B=nc_{0}r_{+}^{2n-4}\left[ c_{0}c_{2}\left( n-1\right)
r_{+}^{2}+c_{1}r_{+}^{3}+\left( n-1\right) \left( n-2\right)
c_{0}^{2}\left\{ r_{+}c_{3}+c_{0}c_{4}\left( n-3\right) \right\} \right] .
\end{equation}

Regarding positive $c_{i}$ leads to positive $B$ and therefore, we should
restrict the parameters to obtain positive $A$. The limitation reduces to
the following inequality
\begin{equation*}
q^{2}l^{2}-2knl^{2}r_{+}^{2n-5}-2n\left( n+1\right) r_{+}^{2n}>0,
\end{equation*}%
where for the flat boundary ($k=0$), one finds
\begin{equation*}
r_{+}<\left( \frac{q^{2}l^{2}}{2n\left( n+1\right) }\right) ^{\frac{1}{2n}}.
\end{equation*}

It is worthwhile to mention that in the absence of massive parameter ($m=0$%
), the obtained solutions will reduce to Reissner--Nordstr\"{o}m
black holes and these black holes only enjoy bound point. Here, in
the presence of massive gravity ($m\neq 0$), there is also only
one root for heat capacity, which is a function of the massive
parameter. It is notable that massive coefficients ($c_{i}$'s),
are only presented in denominator of the obtained critical value.
Therefore, in case of $c_{i}>1$, the effects of these coefficients
on the critical value of massive parameter, are of decreasing
ones. Whereas for the case of $0<c_{i}<1$, their effects are in
favor of increasing the value of the critical massive parameter.
In other words, they increase the value of $m_{c}$.

Next, as for the divergence points of the heat capacity, hence
second order phase transitions, we find following critical
relation for the massive parameter
\begin{equation}
m_{i}\equiv \left. m\right\vert _{C_{Q}\longrightarrow \infty }=\frac{1%
}{2lc_{0}}\sqrt{\frac{C}{D}},  \label{minf}
\end{equation}%
in which%
\begin{equation}
C=6\left[ \left( 2n-1\right) q^{2}l^{2}+2n\left( n+1\right) r_{+}^{2n}\right]
r_{+}-12kl^{2}n\left( n-1\right) r_{+}^{2n-2},
\end{equation}%
\begin{equation}
D=3n\left( n-1\right) r_{+}^{2n-4}\left[ r_{+}^{2}c_{2}+c_{0}\left(
n-2\right) \left\{ 2r_{+}c_{3}+3c_{0}c_{4}\left( n-3\right) \right\} \right]
.
\end{equation}

It is evident that in this case, similar to the case of heat capacity's
root, due to structure of the square root functions, there will be some
restrictions. Interestingly, in this case for $k=0$ and $k=-1$ and due to
our interest only in positive values of massive coefficients, $C$ and $D$
are always positive and there is no restriction. Whereas for the case of
spherical symmetric, there is only one restriction which is in the following
form%
\begin{equation}
\left( 2n-1\right) q^{2}l^{2}+2n\left( n+1\right) r_{+}^{2n}-2l^{2}n\left(
n-1\right) r_{+}^{2n-3}>0,
\end{equation}%
in which due to its complexity it was not possible to find an analytical
restriction with respect to horizon radius. Another interesting property of
the obtained equation for the divergence points of the heat capacity is the
independency of the critical massive parameter of the $c_{1}$ coefficient.
In other words, critical value of the massive parameter in case of
divergencies of the heat capacity is independent of the variation of $c_{1}$%
. Therefore, there is no contribution of the $c_{1}$ to critical
value of the massive parameter. It is worthwhile to mention that,
in this case similar to root of the heat capacity, the presences
of the massive coefficients are only observed in denominator of
$m_{i}$. Therefore, similar effects for variation of the massive
coefficients in case of root of heat capacity will be observed in
this case too.

\begin{figure}[tbp]
$%
\begin{array}{cc}
\epsfxsize=8cm \epsffile{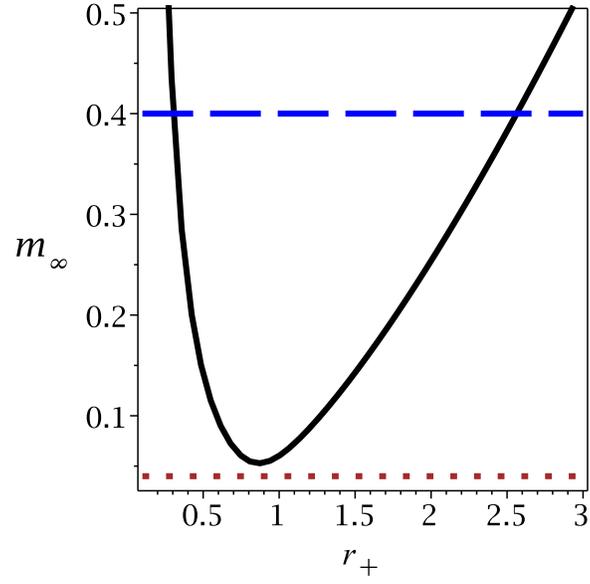} &
\end{array}
$%
\caption{$m_{i}$ (continues line) versus $r_{+}$ for $l=q=1$, $%
c_{0}=c_{2}=c_{3}=c_{4}=2$, $n=4$, and $k=1$. $m_{i}=0.04$ (dotted
line) $m_{i}=0.4$ (dashed line) } \label{Fig1.1}
\end{figure}


\begin{figure}[tbp]
$%
\begin{array}{cc}
\epsfxsize=6cm \epsffile{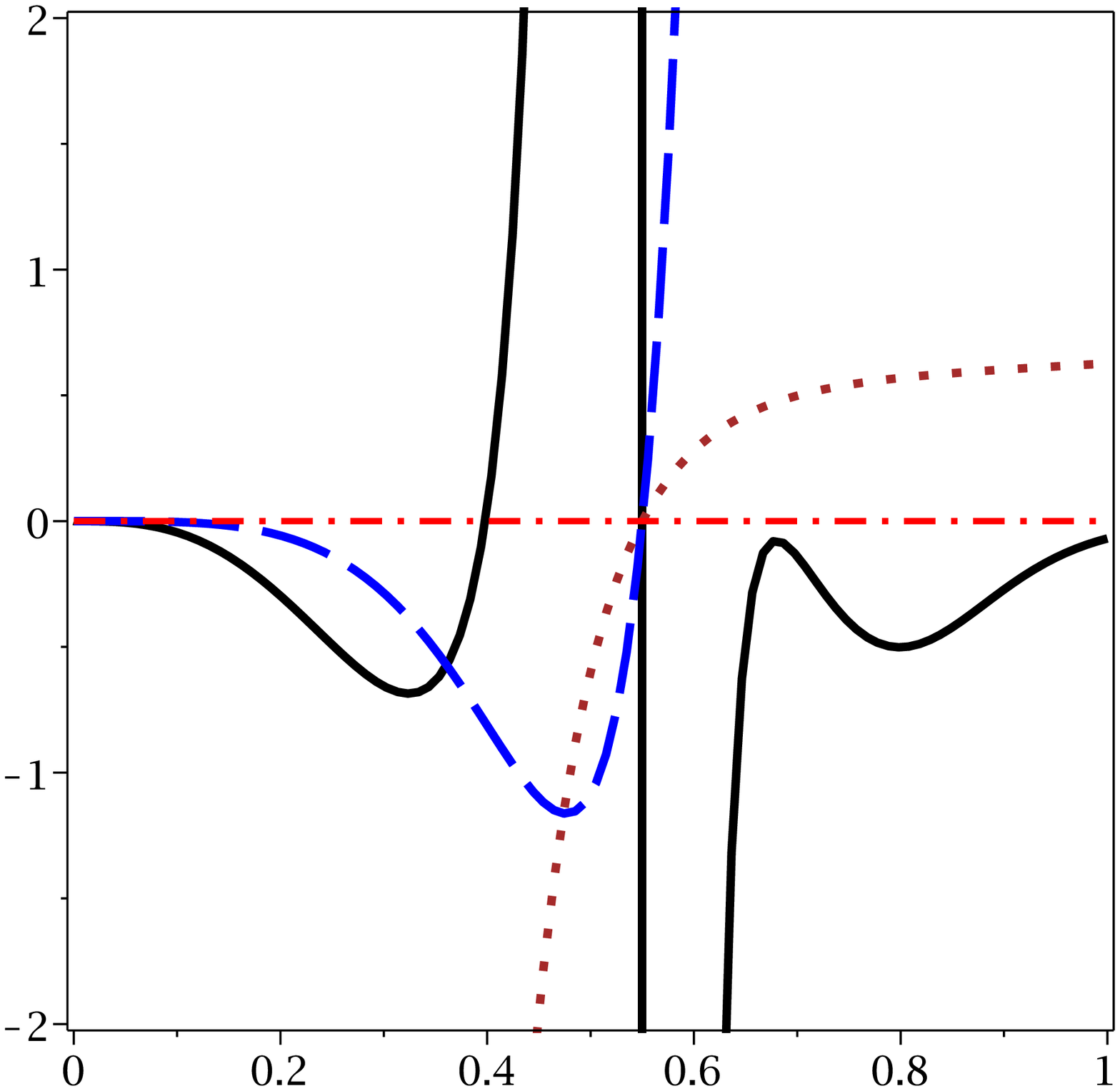} & \epsfxsize=6cm \epsffile{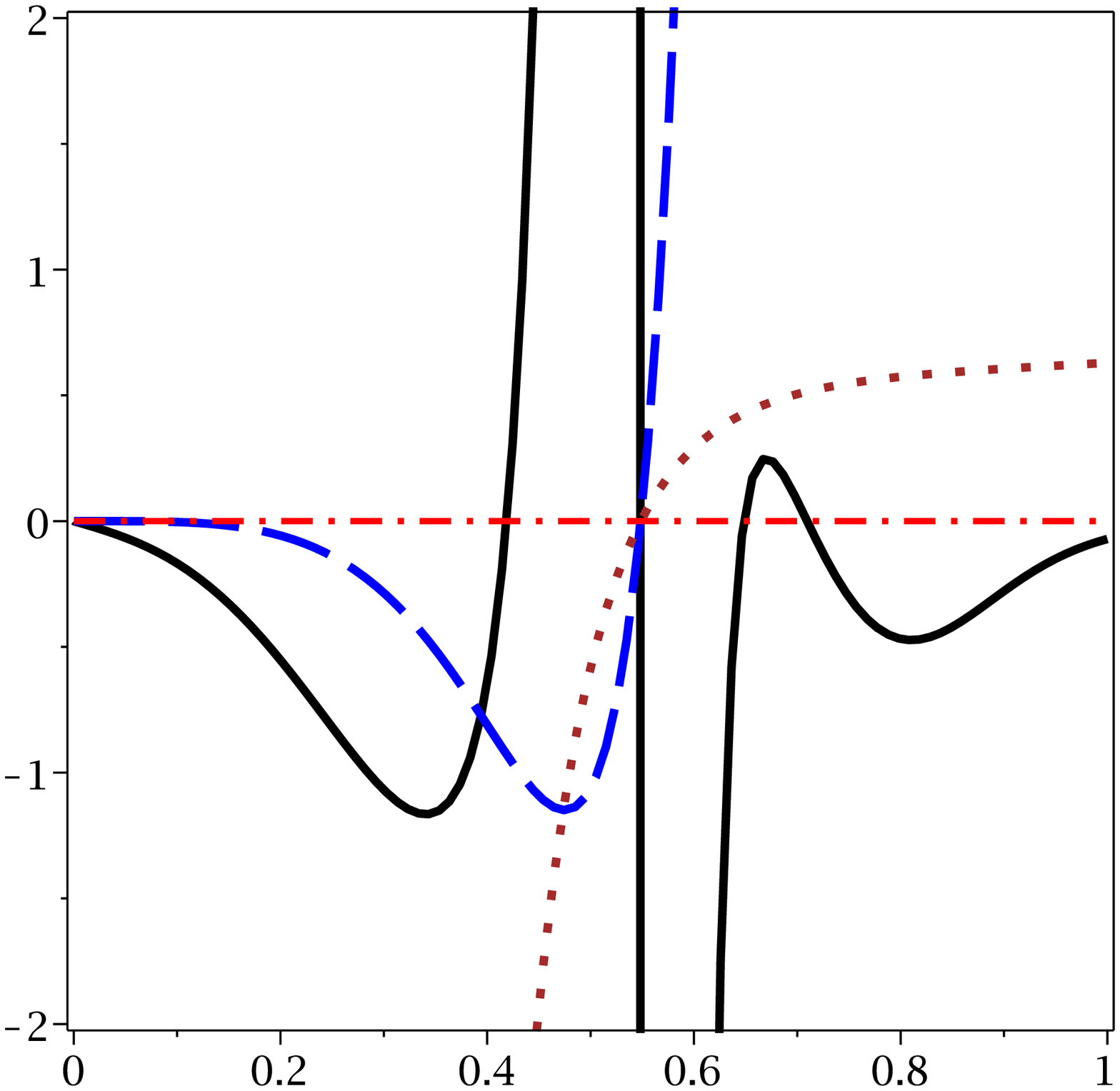} \\
\epsfxsize=6cm \epsffile{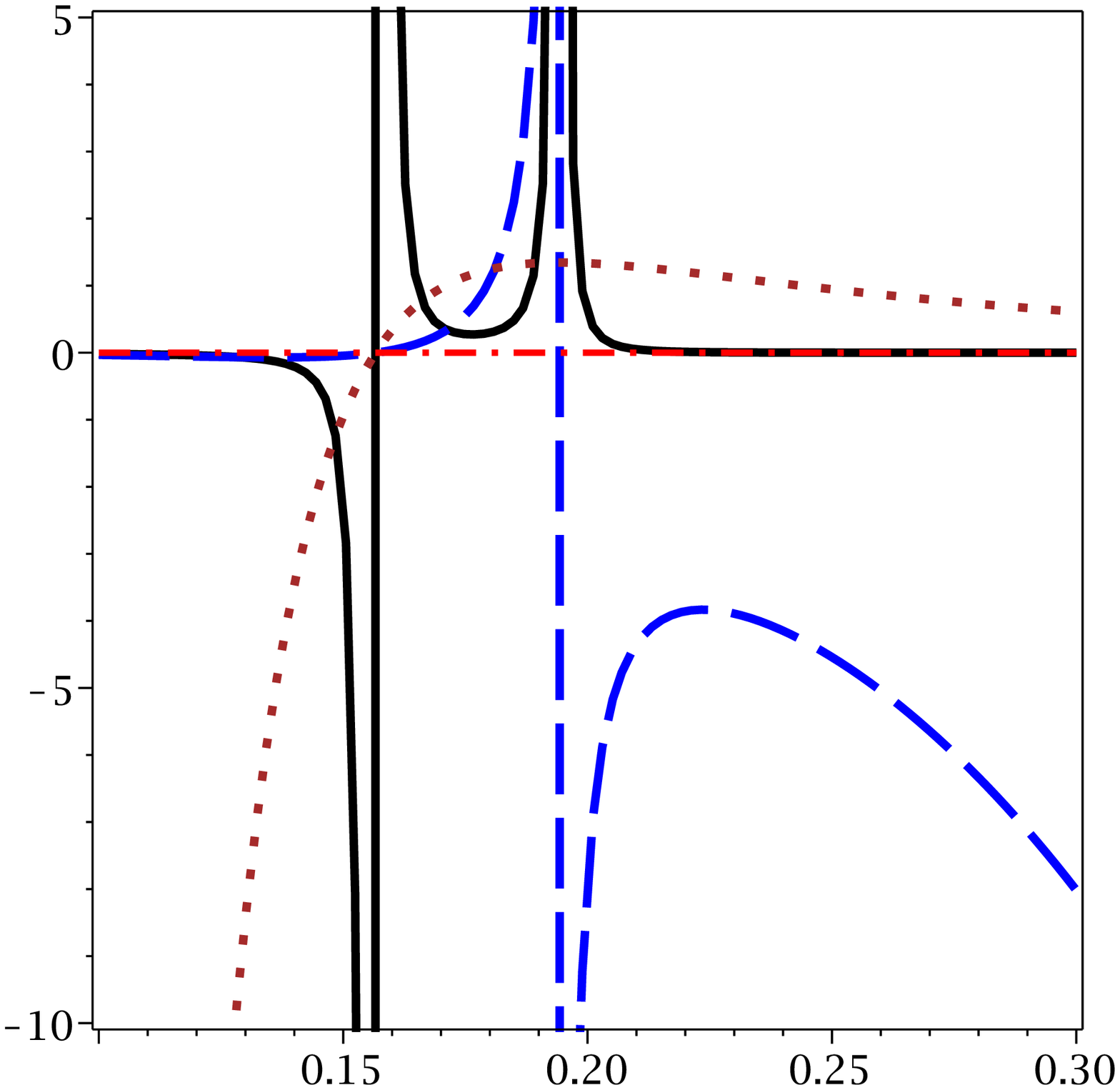} & \epsfxsize=6cm \epsffile{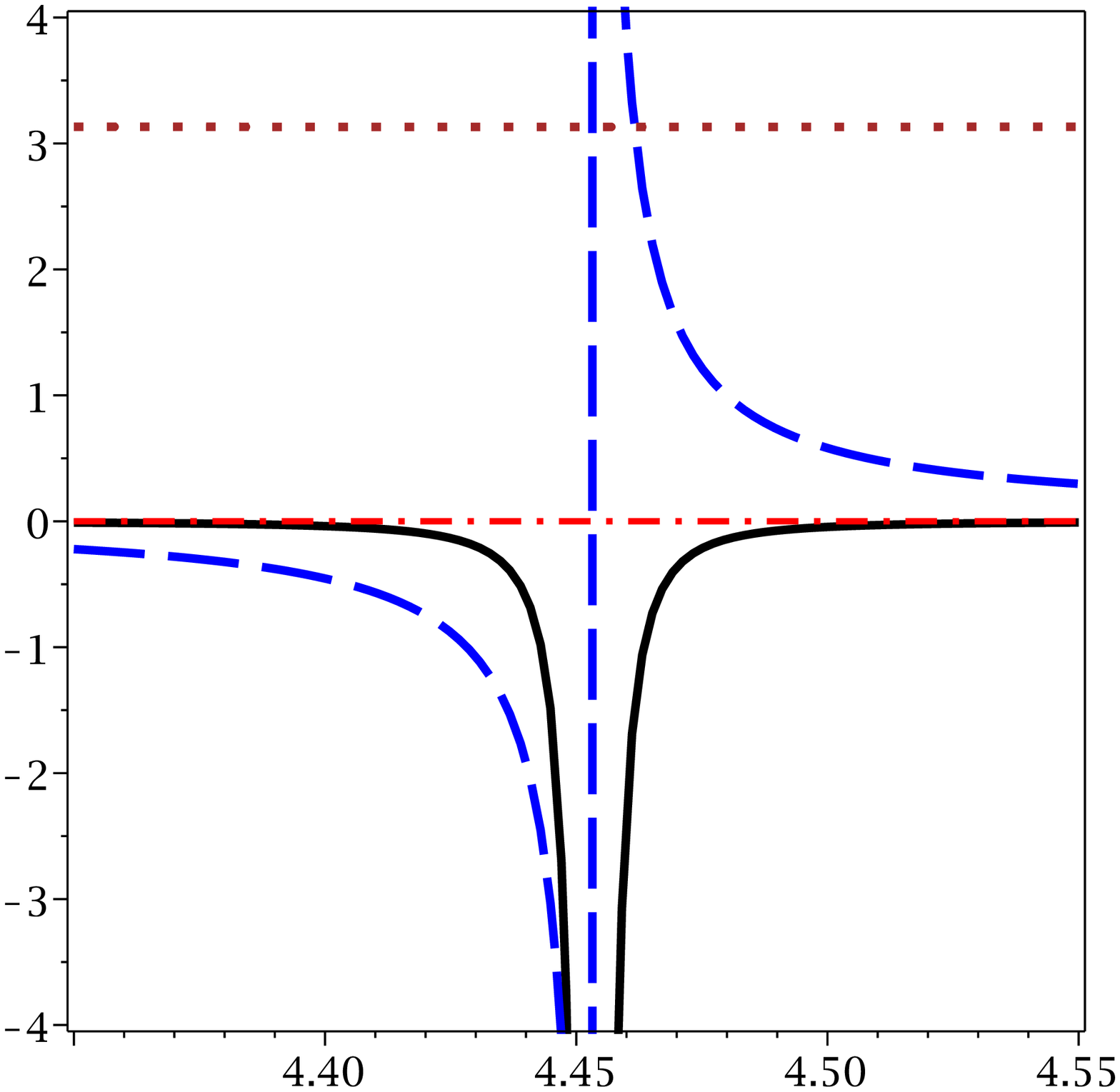}%
\end{array}
$%
\caption{$\mathcal{R}$ (continuous line), $C_{Q}$ (dashed line) and $T$
(dotted line) versus $r_{+}$ for $l=q=1$, $c_{0}=c_{1}=c_{2}=c_{3}=c_{4}=2$,
$n=4$, and $k=1$; $m=0$ (up - left), $m=0.01$ (up - right) and $m=1$ (down
panels with different scales).}
\label{Fig1}
\end{figure}


\begin{figure}[tbp]
$%
\begin{array}{cc}
\epsfxsize=6cm \epsffile{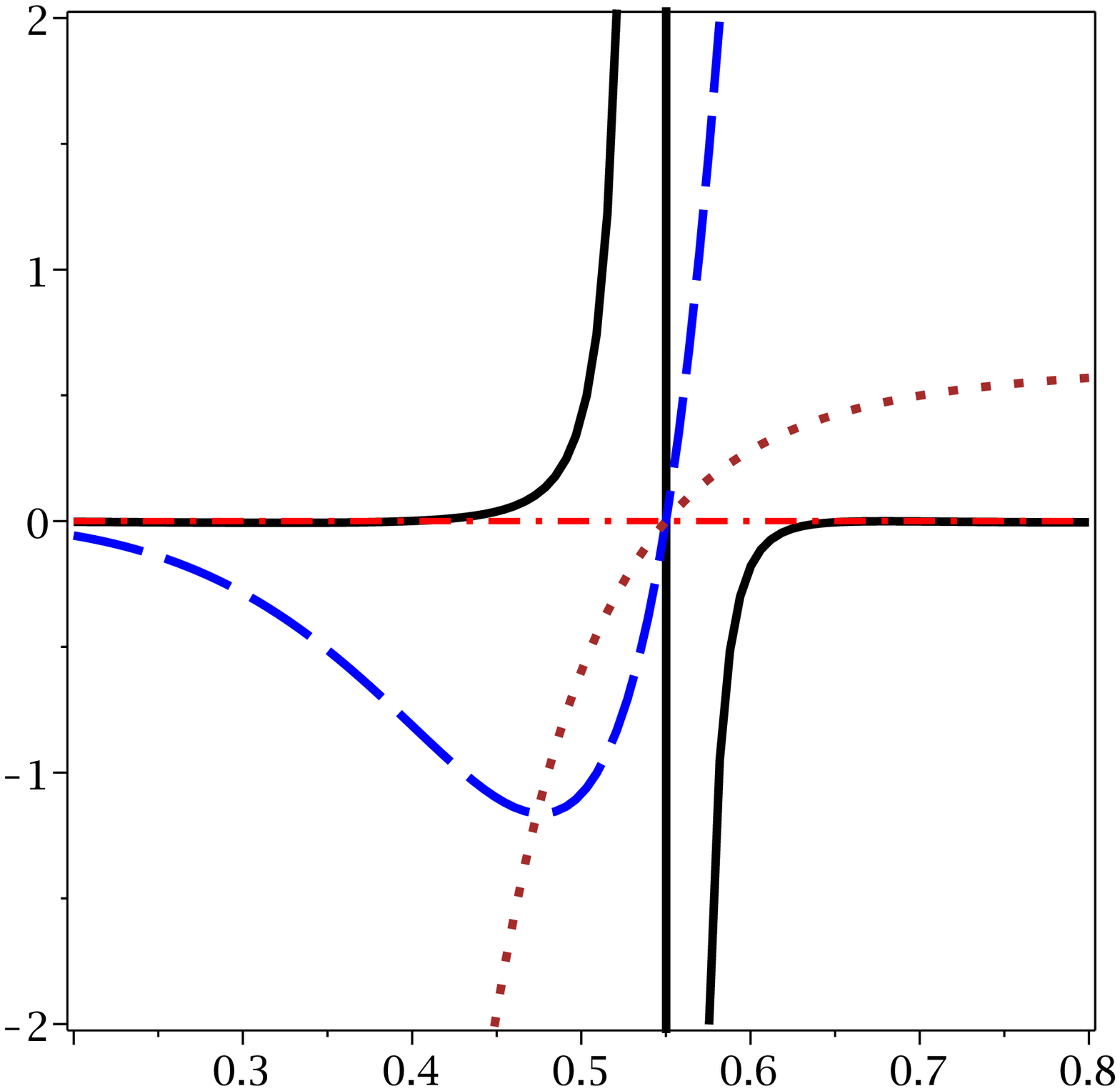} & \epsfxsize=6cm \epsffile{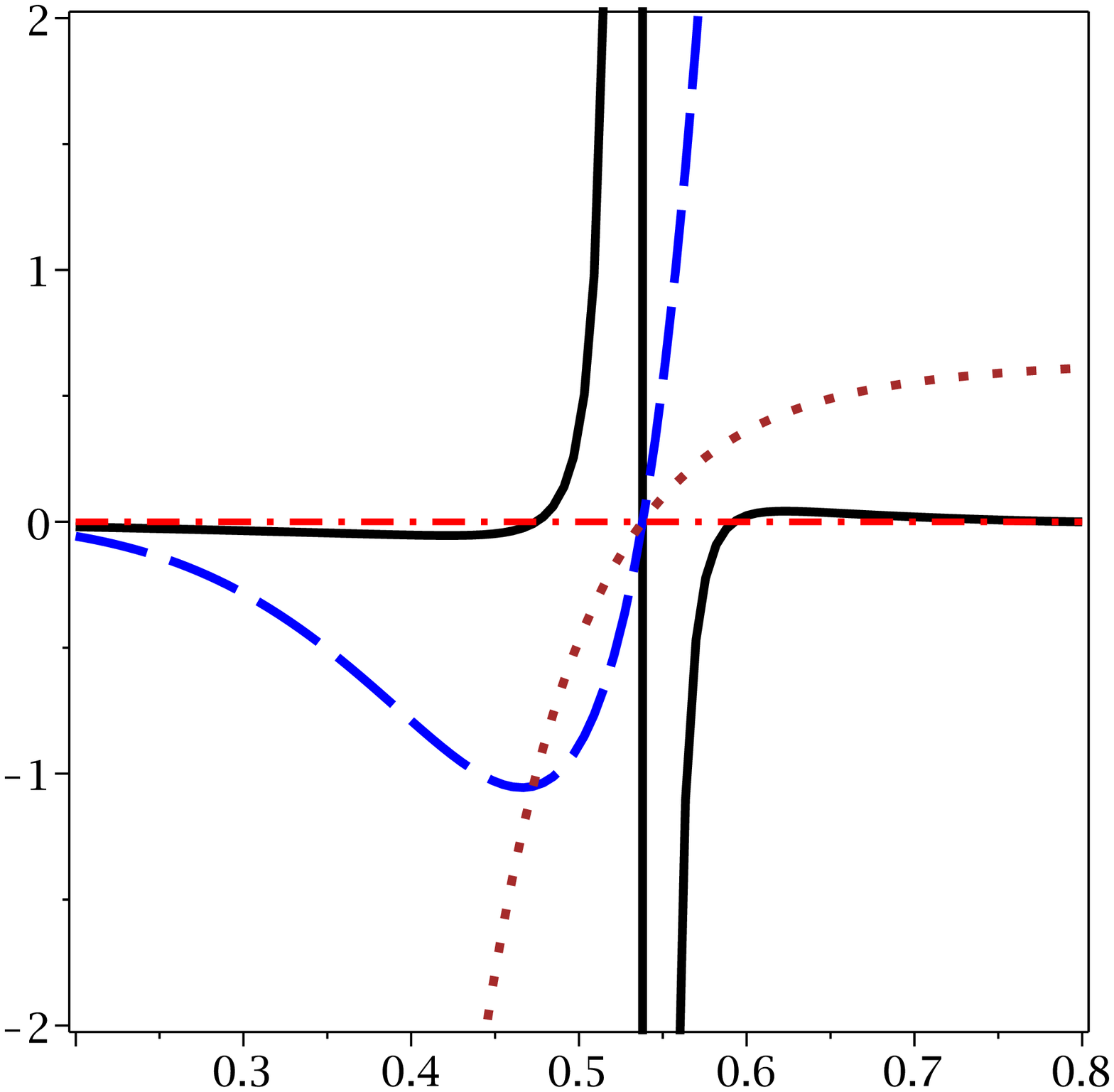} \\
\epsfxsize=6cm \epsffile{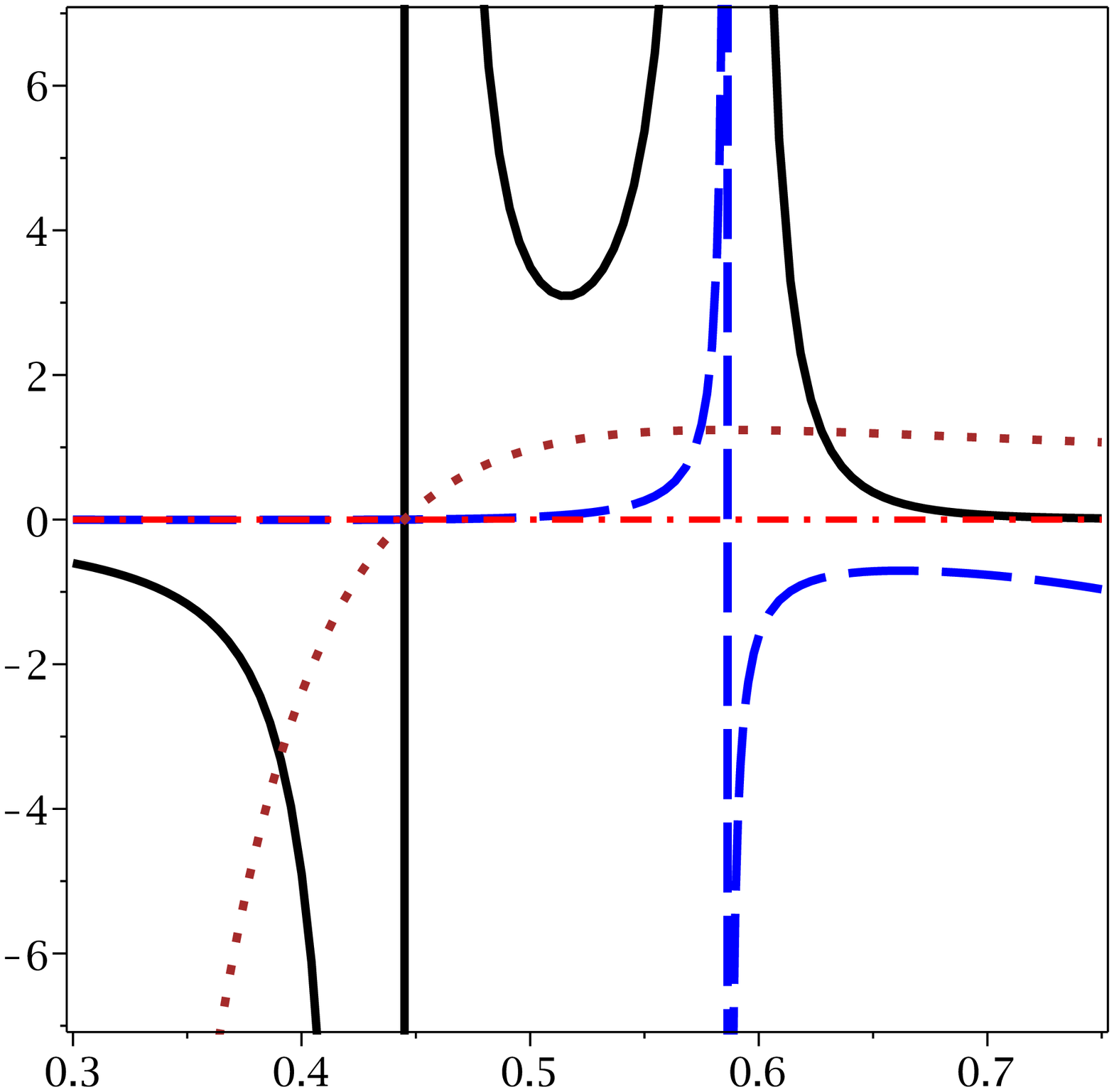} & \epsfxsize=6cm \epsffile{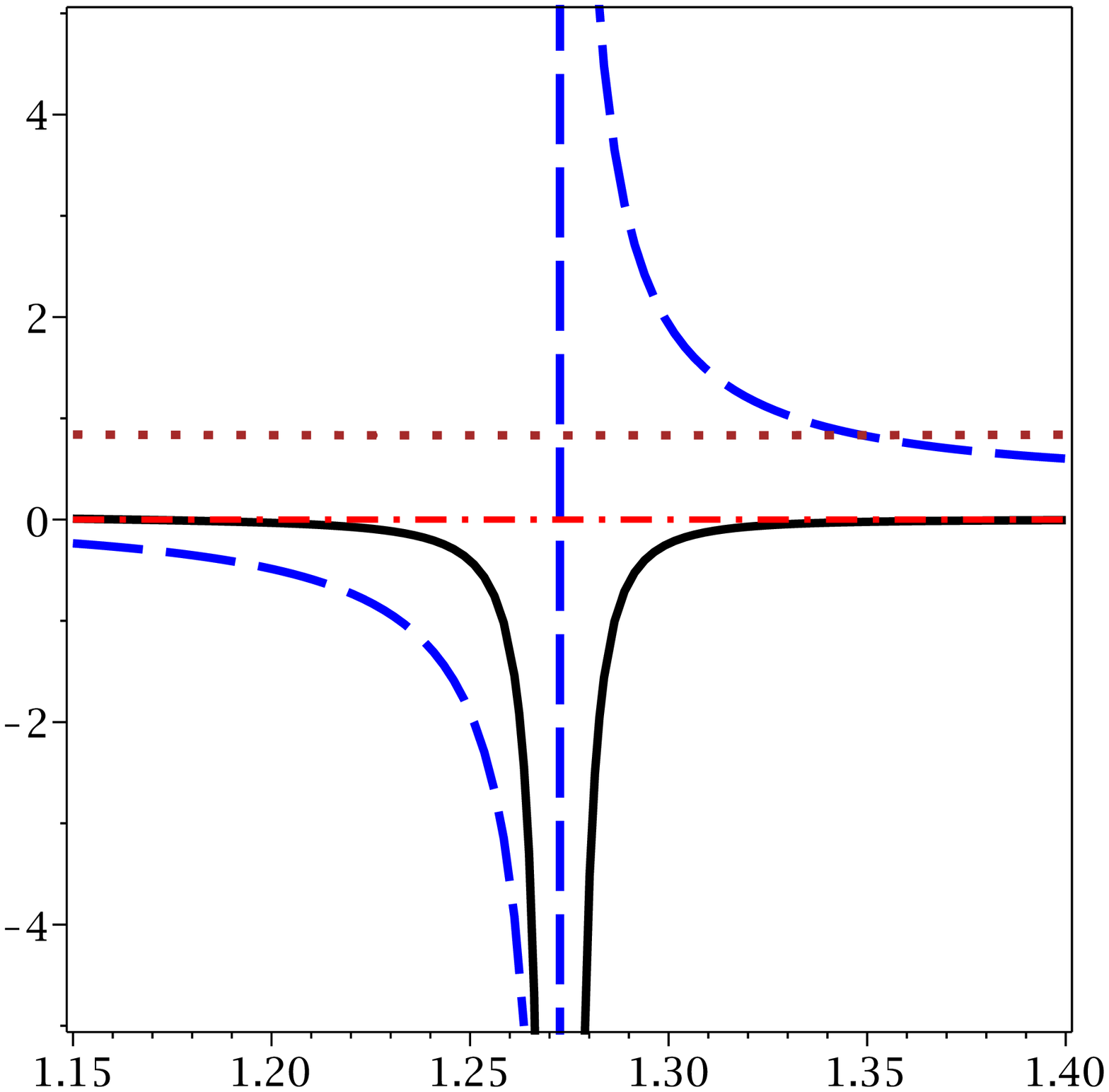}%
\end{array}
$%
\caption{$\mathcal{R}$ (continuous line), $C_{Q}$ (dashed line) and $T$
(dotted line) versus $r_{+}$ for $l=q=1$, $c_{1}=c_{2}=c_{3}=c_{4}=2$, $n=4$%
, $m=0.1$ and $k=1$; $c_{0}=0$ (up - left), $c_{0}=1$ (up - right) and $%
c_{0}=2$ (down panels with different scales).}
\label{Fig2}
\end{figure}


\begin{figure}[tbp]
$%
\begin{array}{ccc}
\epsfxsize=6cm \epsffile{m011.eps} & \epsfxsize=6cm \epsffile{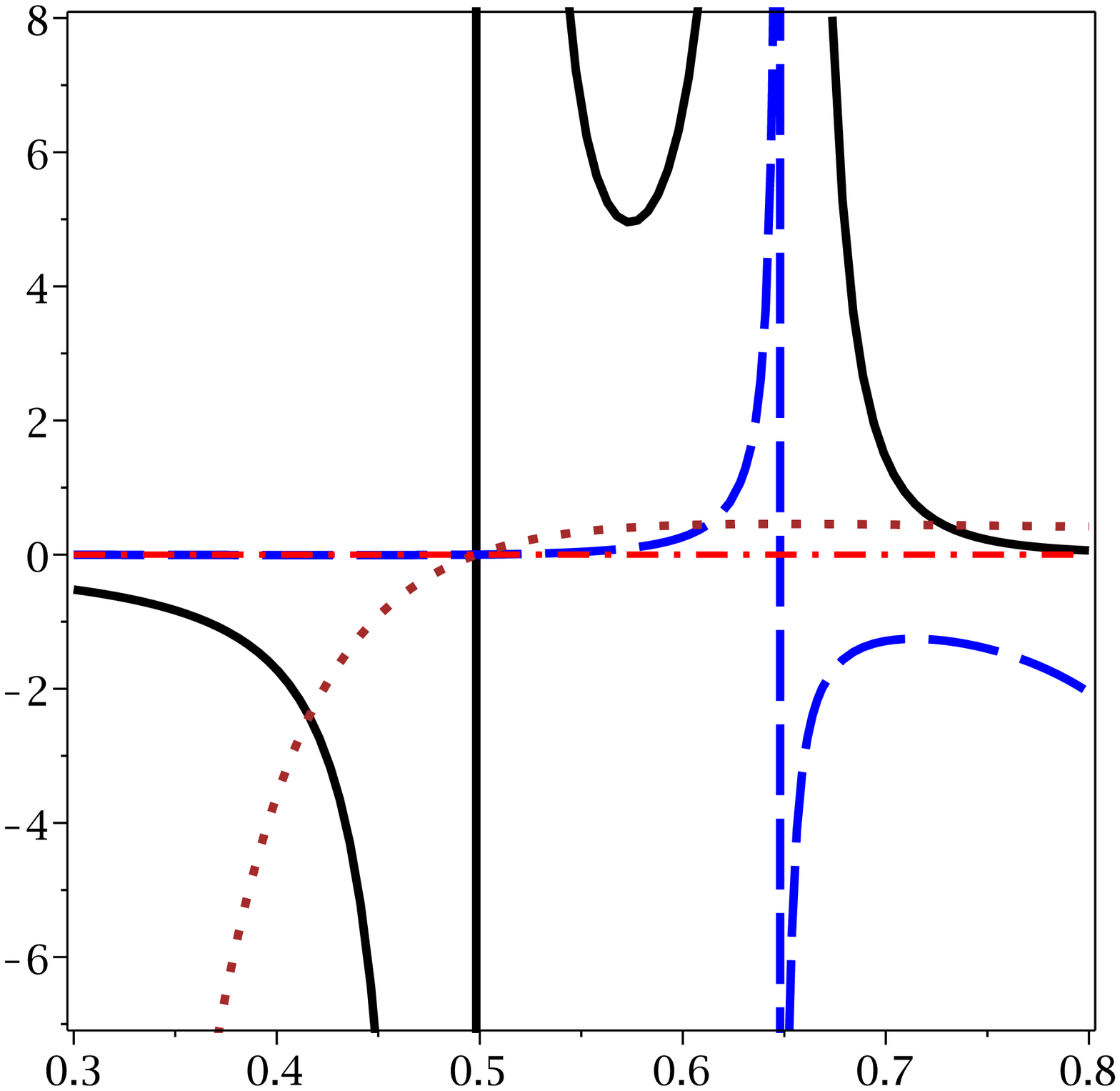} & %
\epsfxsize=6cm \epsffile{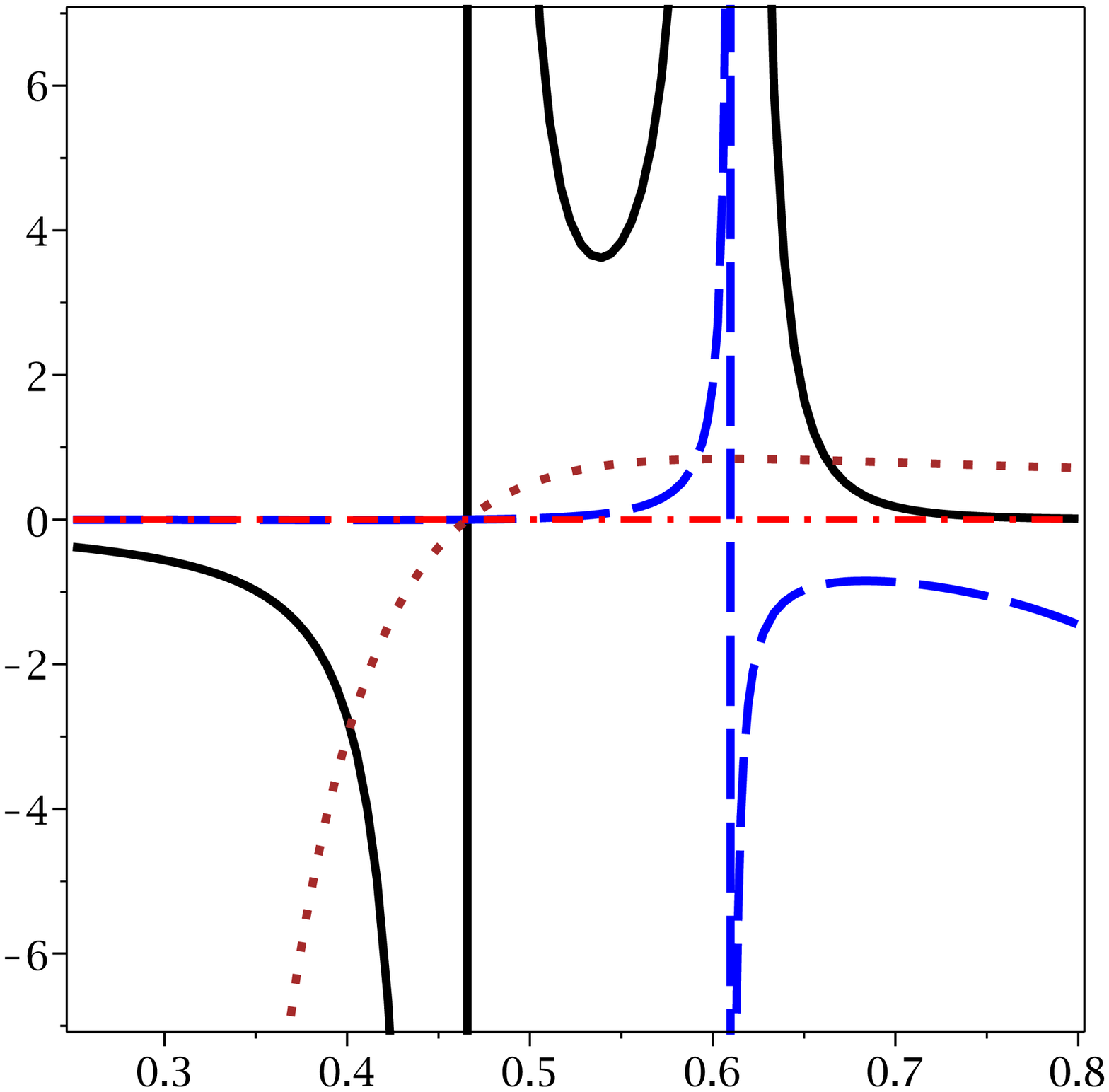} \\
\epsfxsize=6cm \epsffile{m012.eps} & \epsfxsize=6cm \epsffile{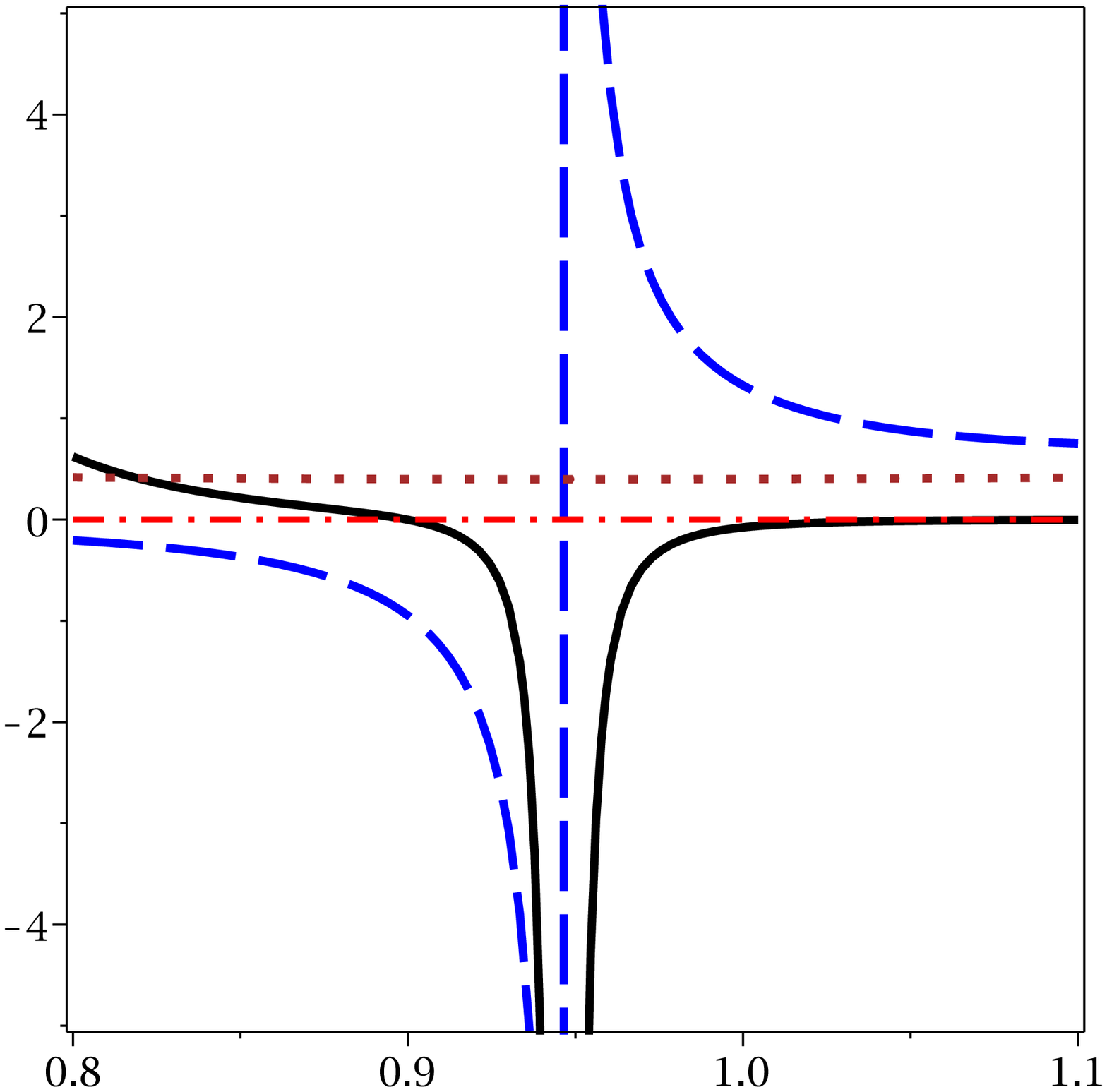} & %
\epsfxsize=6cm \epsffile{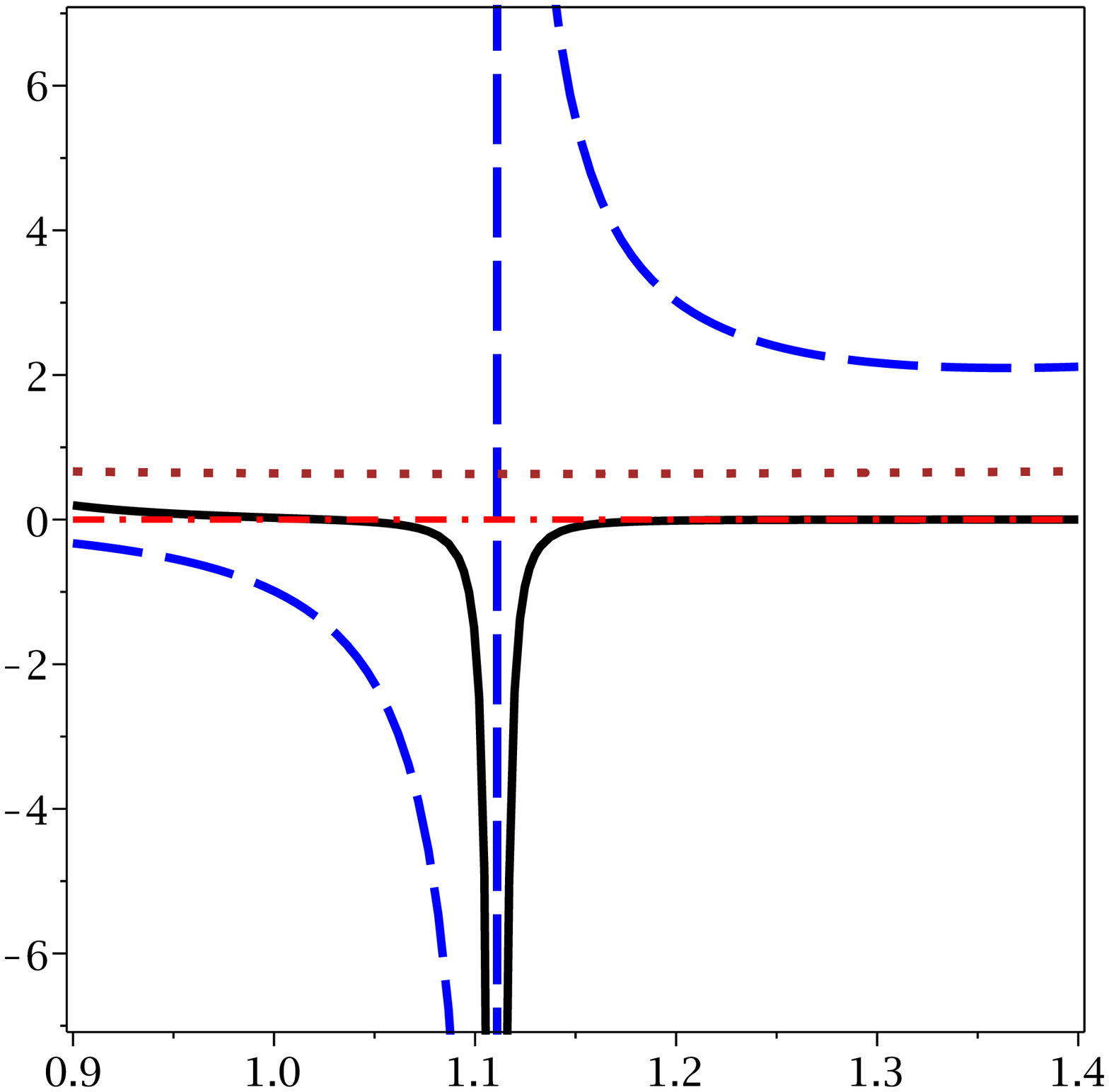}%
\end{array}
$%
\caption{$\mathcal{R}$ (continuous line), $C_{Q}$ (dashed line) and $T$
(dotted line) versus $r_{+}$ for $l=q=1$, $c_{0}=c_{1}=c_{2}=c_{3}=c_{4}=2$,
$n=4$, and $m=0.1$; for different scale: $k=1$ (left panels up and down), $%
k=-1$ (middle panels up and down) and $k=0$ (right panels up and down).}
\label{Fig3}
\end{figure}


\begin{figure}[tbp]
$%
\begin{array}{cc}
\epsfxsize=7cm \epsffile{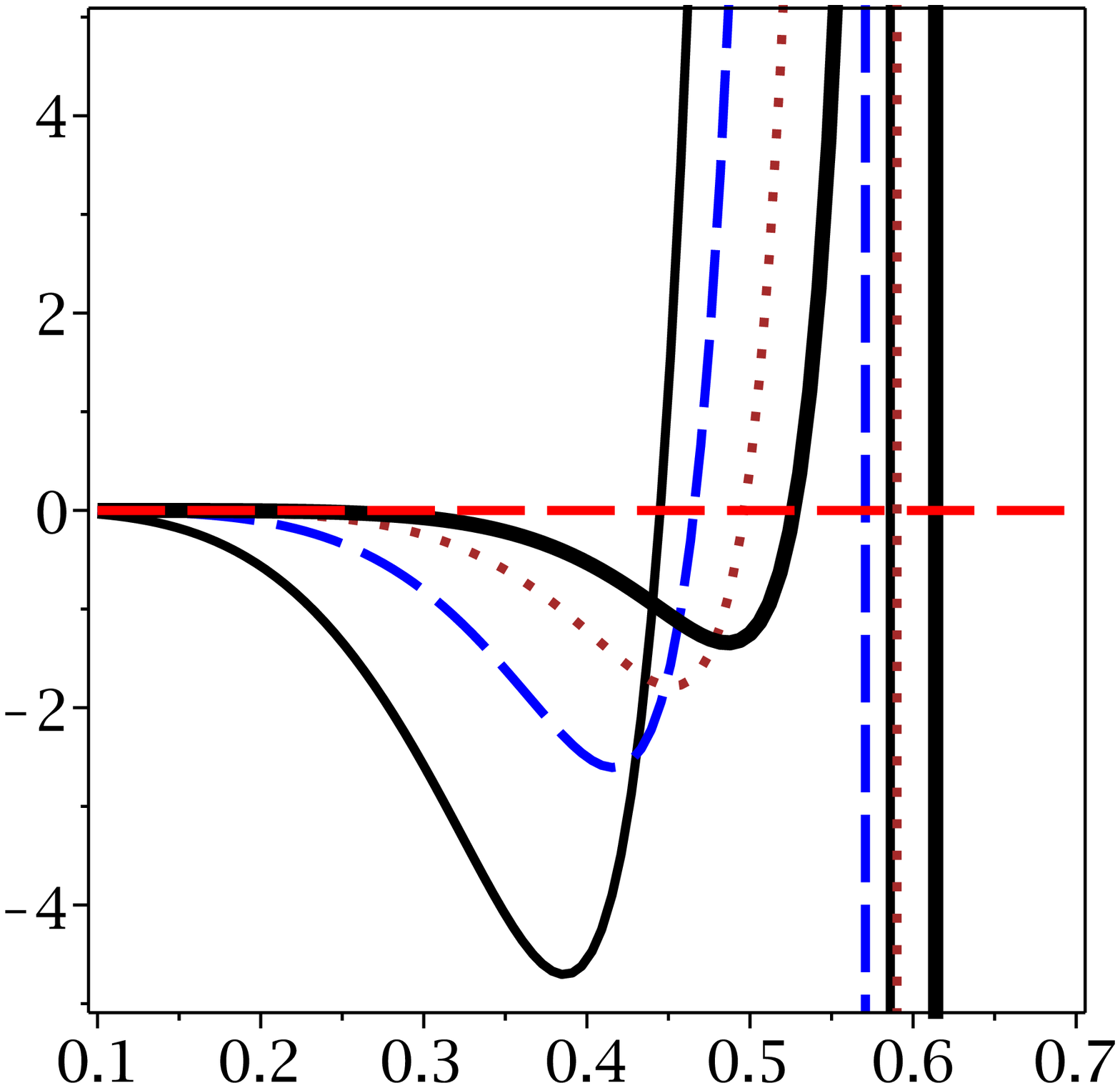} & \epsfxsize=7cm %
\epsffile{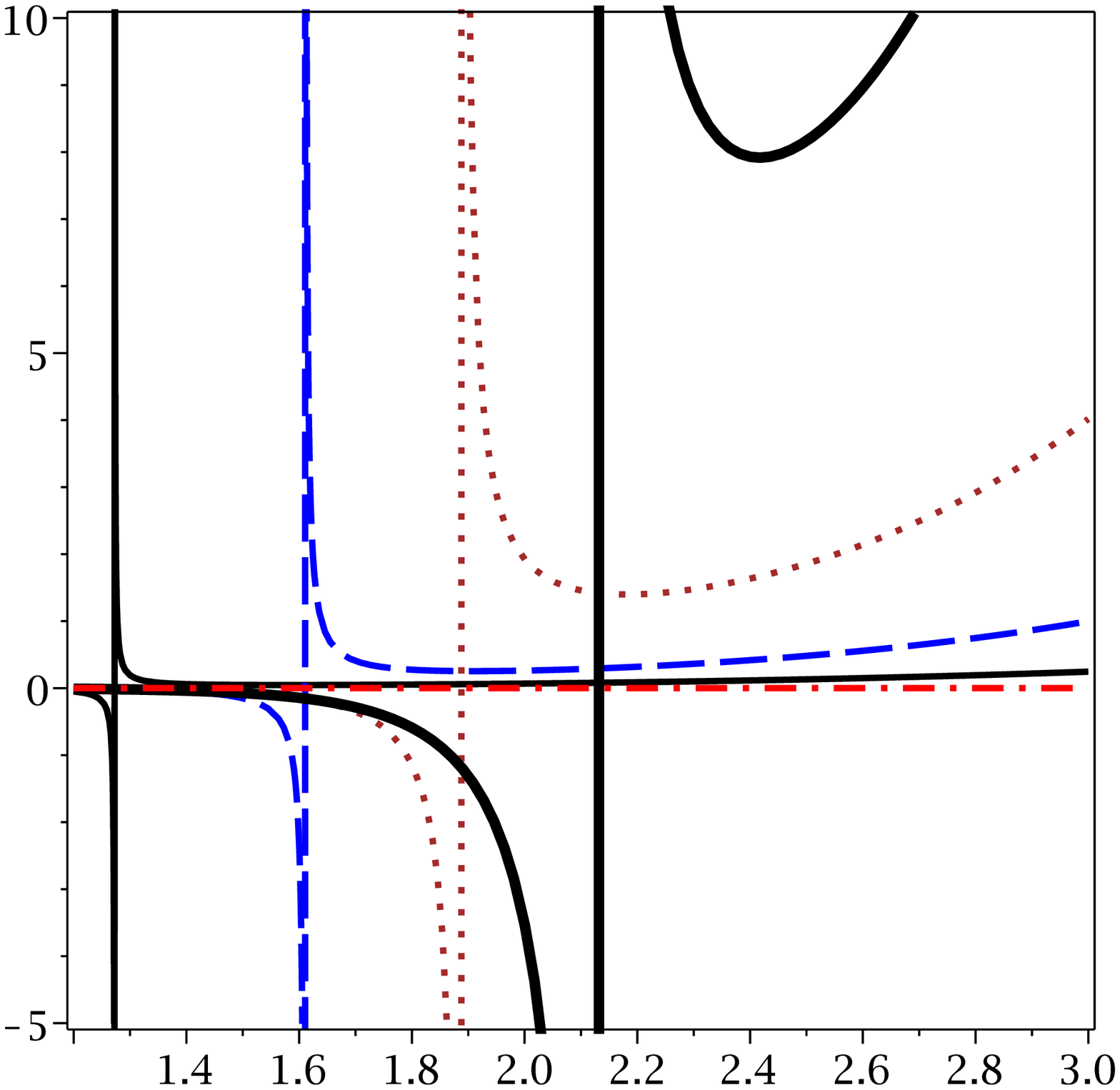}%
\end{array}
$%
\caption{ $C_{Q}$ versus $r_{+}$ for $l=q=1$, $%
c_{0}=c_{1}=c_{2}=c_{3}=c_{4}=2$, $k=1$, and $m=0.1$. \newline
for different scales: $n=4$ (continues line), $n=5$ (dashed line), $n=6$
(dotted line) and $n=7$ (bold continues line).}
\label{Fig4}
\end{figure}


\begin{figure}[tbp]
$%
\begin{array}{cc}
\epsfxsize=7cm \epsffile{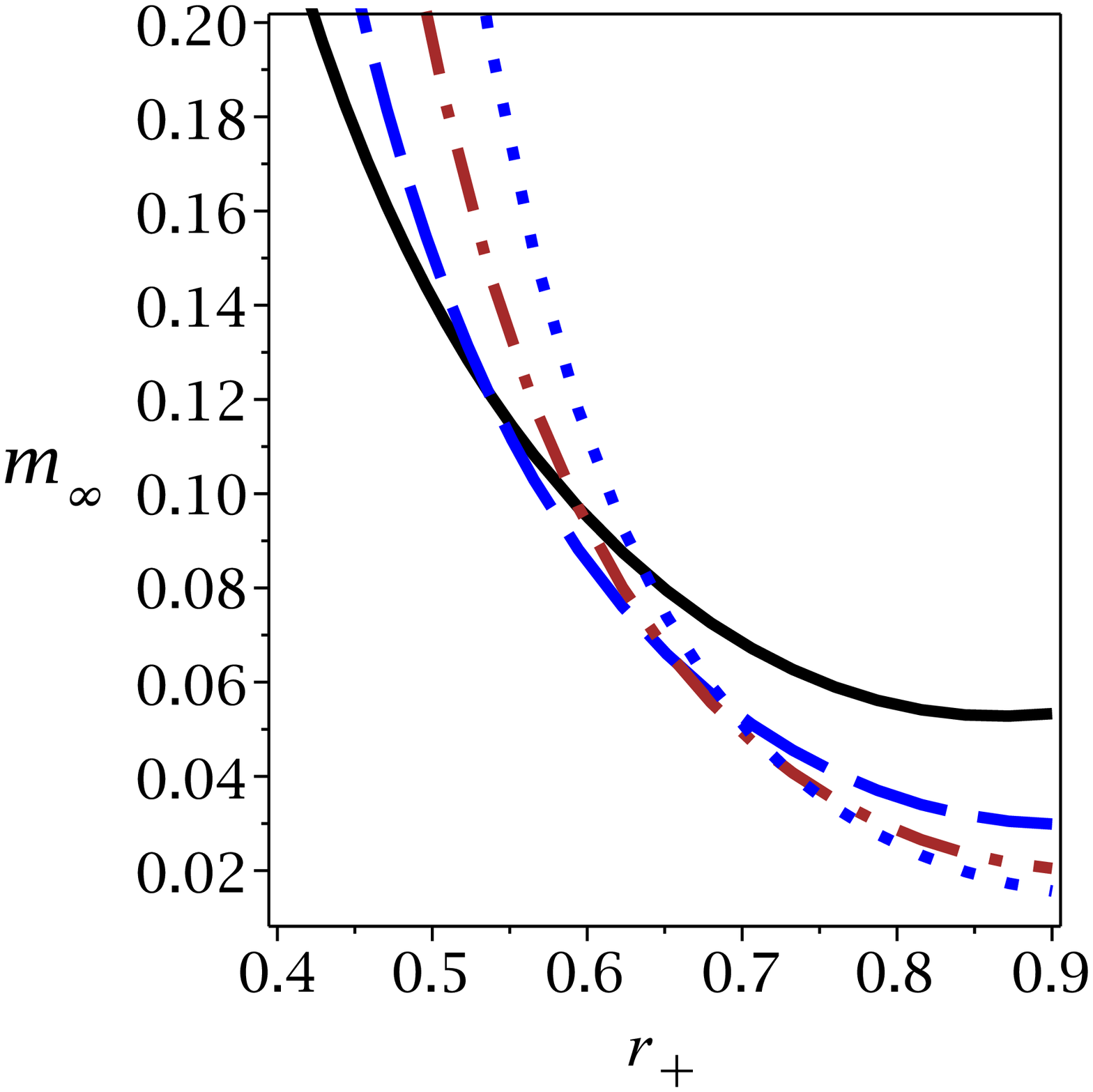} & \epsfxsize=7cm %
\epsffile{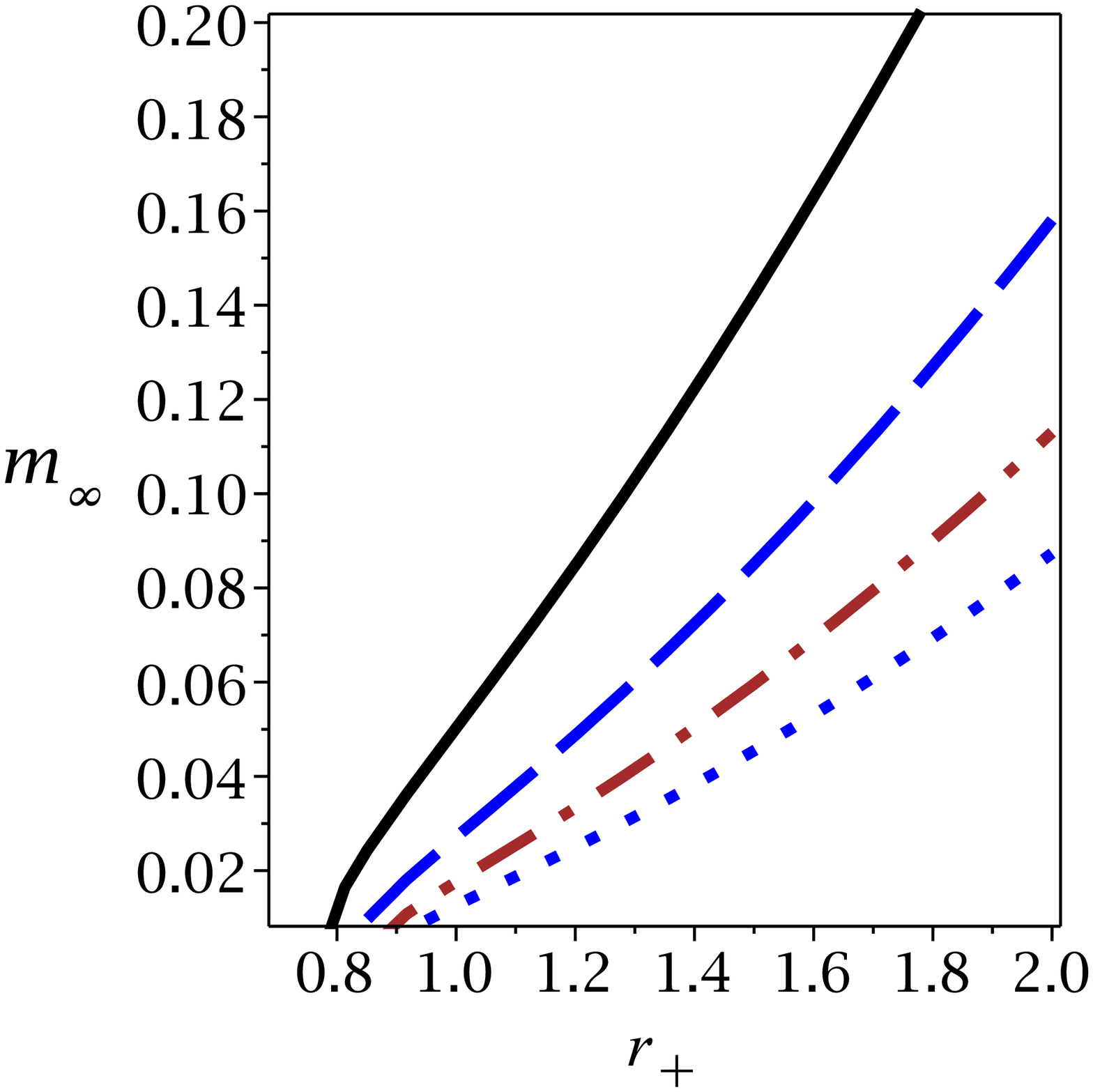}%
\end{array}
$%
\caption{$m_{i}$ (continues line) versus $r_{+}$ for $l=1$, $%
c_{0}=c_{1}=c_{2}=c_{3}=c_{4}=2$, $k=1$, $n=4$ (continues line), $n=5$
(dashed line), $n=6$ (dashed-dotted line) and $n=7$ (dotted line); $q=1$
(left panel) and $q=0$ (right panel) }
\label{Fig1.2}
\end{figure}


In addition, it must be pointed out that the obtained value of
massive parameter in Eq. (\ref{minf}) is not an injective
function. Plotting $m_{i}$
with respect to horizon radius, one finds that there is a value of $%
m_{i}$, say $m_{i_{C}}$ in which for $m_{i}<m_{i_{C}} $, there
will be no critical horizon radius available. In other words, in
this region there is no divergence point for the heat capacity. In
case of $m_{i}=m_{i_{C}}$, there will be only one divergence point
for heat capacity and finally for the case of $m_{i}>m_{i_{C}}$,
there will be two horizon radii for any critical value of the
massive parameter. This property of the obtained equation will be
shown in the plotted graphs for the heat capacity. Here, in order
to elaborate mentioned behavior for Eq. (\ref{minf}), we have
plotted Fig. \ref{Fig1.1}.

Next, by employing the Eqs. (\ref{temp}), (\ref{heatcap}) and
(\ref{HPEM}), we plot Figs. \ref{Fig1}-\ref{Fig4} to study
thermodynamic and geometrical thermodynamic behavior of these
black holes.

It is evident that for the small values of massive parameter,
there is no second order phase transition (Fig. \ref{Fig1} up).
Whereas for large values of this parameter, there is a bound point
and a second order phase transition (Fig. \ref{Fig1} down). It is
worthwhile to mention that this behavior could be observed by
studying Fig. \ref{Fig1.1}. Also, according to the plotted graph
for $m_{i} $ (Fig. \ref{Fig1.1}), the smaller (larger) divergence
point of the heat capacity is a decreasing (an increasing)
function of $m$.
This behavior is also evident from plotted graphs for variation of $m$ (Fig. %
\ref{Fig1}).

Besides, due to coupling of different powers of $c_{0}$ with massive
parameter, we have studied the effects of variation of this coefficient.
Interestingly, similar to behavior of the massive parameter, for small
values of $c_{0},$ only a bound point is observed (Fig. \ref{Fig2} up). It
is seen that only for sufficiently large values of this coefficient, two
second order phase transitions and a bound point are obtained (Fig. \ref%
{Fig2} down).

Our next effects of interest are due to variation of topology of the
solutions. If we denote the indexes $F$, $H$ and $S$ as representing flat ($%
k=0$), hyperbolic ($k=-1$) and spherical horizons respectively, and also
root of heat capacity with $r_{0}$ and two divergence points of the heat
capacity with $r_{Div1}$ (smaller one) and $r_{Div2}$ (larger one), we can
find following results. For the root of the heat capacity, we have $%
r_{0-S}<r_{0-F}<r_{0-H}$. In other words, the highest value for
the root of heat capacity belongs to the solutions with hyperbolic
topology. In addition, one can find that
$r_{Div1-S}<r_{Div1-H}<r_{Div1-F}$ which indicates that the
highest value of smaller divergence point belongs to flat horizon
(Fig. \ref{Fig3} up). Finally, for the larger divergence point,
one can find that $r_{Div1-H}<r_{Div1-F}<r_{Div1-S}$ which means
that in this case the highest value of larger divergence point
belongs to spherical symmetric black holes (Fig. \ref{Fig3} down).

Finally, we are in a position to study the effects of dimensions
on thermodynamical behavior of the system. In plotted graph, we
only present the heat capacity diagrams for various dimensions.
Evidently the root of the heat capacity is only an increasing
function of dimensions (Fig. \ref{Fig4} left). As for divergence
point of the heat capacity, the larger one is also an increasing
function of dimensions (Fig. \ref{Fig4} right). But regarding the
smaller divergence point, interestingly, an anomaly is observed in
case of $n=5$. In other words, for increasing value of dimensions,
the smaller divergence point increases except for the case of
$n=5$ which its smaller divergence point is smaller than the case
of $n=4$. In order to find the reason for this behavior, once
again we plot Fig. \ref{Fig1.2} (left) for different dimensions.
Evidently, for specific small values of $m$, the behavior of the
system is with order, in which increment in dimensions leads to
increase in value of the smaller divergence point of the heat
capacity. On the other hand, for special values of $m$ the smaller
divergence points of different dimensions will be the same. In
other words, there are some values of $m$ in which smaller
divergence point of the heat capacity will be the same for $n=4$
and $n=5$ or other dimensions. This behavior of the system is
solely because of the structure of $m_{i} $, and it is the
characteristic behavior of the system with this configuration (for
this type of massive gravity). It is worthwhile to mention that
increasing the value of dimensions will decrease the region in
which there is no second order phase transition. In other words,
the minimum of the $m_{i}$ is a decreasing function of dimensions.
Interestingly, for the case of vanishing electric charge (Fig.
\ref{Fig1.2} right), the mentioned behavior will not be seen. In
this case, there is only one second order phase transition for
these black holes which specifies the role of the electric charge
in thermodynamical behavior of the system.

Now, we focus to study thermal stability of the solutions. The
stability of the solutions, hence, thermodynamical behavior of
these black holes are presented by the bound point and number of
the second order phase transitions. For example, in case of only
one bound point (up panels of Figs. \ref{Fig2} and \ref{Fig3}),
there is a region in which both temperature and heat capacity are
negative. Therefore, this region is presenting the non-physical
unstable state for black holes. After passing the bound point, the
system acquires a stable physical state. For the cases of two
second order phase transitions and one bound point (down panels of
Figs. \ref{Fig2} and \ref{Fig3}), there are several changes in the
sign of heat capacity. By using naming that was used before, one
can find that for the case of $r_{+}<r_{0}$, system is in
nonphysical and unstable phase. Whereas in $r_{+}=r_{0}$ system
goes under a transition from nonphysical to physical state and for
$r_{0}<r_{+}<r_{Div1}$, system is
in thermally stable state with positive temperature. As for the $%
r_{Div1}<r_{+}<r_{Div2},$ system is in unstable state and for $r_{Div2}<$ $%
r_{+}$, black holes are thermally stable with positive
temperature. Therefore, due to thermodynamical concept in which
system desires to find
stable state, there are two second order phase transitions. For the case of $%
r_{+}=r_{Div1}+\delta $, in which $\delta $ is so small, the system goes
under a second order phase transition from larger unstable black hole to
smaller stable black hole. In other words, in this phase transition the size
of the black hole decreases. On the other hand, for the case of $%
r_{+}=r_{Div2}-\delta $, black hole in this configuration goes under another
second order phase transition. In this one, black hole acquires larger
horizon radius, hence its size increases.

Next, we study the geometrical thermodynamic behavior of the system. Figs. %
\ref{Fig1} to \ref{Fig4} show that considered metric in the
context of GTs, provided a successful machinery. In other words,
bound point and second order phase transitions for the heat
capacity and the divergence points of Ricci scalar of the
considered metric coincide with each other. Therefore, the
mentioned metric has divergencies in place of bound point as well
as second order phase transition points. On the other hand, the
behavior of the Ricci scalar near bound point and second order
phase transitions is not the same. For the root of the heat
capacity (bound point) the sign of TRS is different for before and
after that point. Whereas, in case of the second order phase
transition, the sign of TRS is fixed. This change in sign enables
one to distinguish bound point from second order phase transition.
Therefore, it is possible to find bound point and phase transition
by applying GTs, without the use of heat capacity diagrams.

\section{critical behavior in extended phase space}

\subsection{critical behavior in extended phase space through usual method}

The van der Waals like behavior of the black holes could be
studied through the use of the analogy between cosmological
constant and thermodynamical pressure. The proportionality between
cosmological constant and pressure is given by

\begin{equation}
P=-\frac{\Lambda }{8\pi }.  \label{P}
\end{equation}

By replacing the cosmological constant with thermodynamical
pressure in temperature, one can obtain the equation of state.
Using this equation of state and inflection point property, one
can extract critical points and study van der Waals like behavior.
For black hole solutions which considered in this paper, the
critical behavior of the system through usual method has been
investigated in Ref. \cite{Xu2015}. Here, we would like to
investigate the critical properties of the black holes through a
new method.

\subsection{critical behavior in extended phase space through new method}

In this section, we will study the critical behavior of these black holes
through the use of the proportionality between pressure and cosmological
constant. The method which we employ to do so, was introduced in Ref. \cite%
{HPE}. By replacing the cosmological constant with its
corresponding thermodynamical pressure in the heat capacity, one
is able to obtain a relation for the pressure by solving the
denominator of heat capacity. The maximum of $P(r_{+})$ is where
the second order phase transition takes place. In this relation,
$P$ does not explicitly depend on the temperature, and therefore,
this relation differs from equation of state, but it contains all
the information regarding phase structure of the black holes. For
the pressures smaller than critical pressure, two horizon radii
for each pressure exist which indicate three phases. While for
pressures larger than critical pressure, no phase transition is
observable.

Using the denominator of heat capacity (\ref{heatcap}) with
(\ref{P}), one can find following relation for the pressure
\begin{eqnarray}
P&=&\frac{3n( n-1)^{2}( n-2)(n-3) c_{0}^{4}m^{2} c_{4}}{16\pi r_{+}^{4}(1+n) }+%
\frac{n( n-1) ^{2}(n-2) c_{0}^{3}{m}^{2}{c}_{{3}}}{{8}\pi r_{+}^{3}(1+n) }+%
\frac{n( n-1)^{2} c_{0}^{2}{m}^{2}{c}_{{2}}}{16\pi r_{+}^{2} (1+n) }-  \nonumber \\
&&\frac{(n-1) (2n-1) {q}^{2}}{32\pi r_{+}^{2n} (1+n) }+\frac{k n (n-1)^{2}}{%
16\pi r_{+}^{2} (1+n) }.  \label{PP}
\end{eqnarray}

\begin{figure}[tbp]
$%
\begin{array}{cc}
\epsfxsize=6cm \epsffile{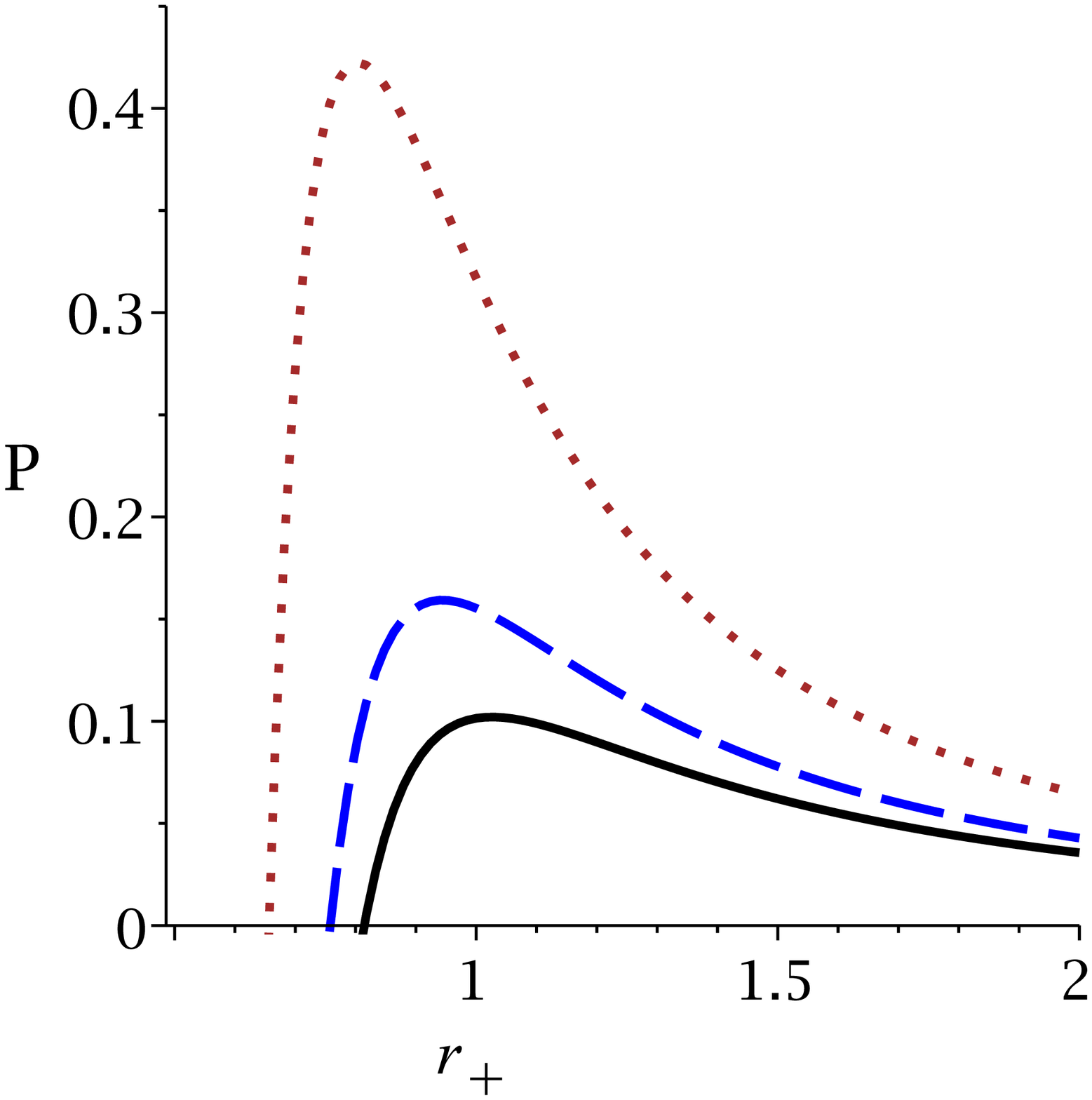} & \epsfxsize=6cm %
\epsffile{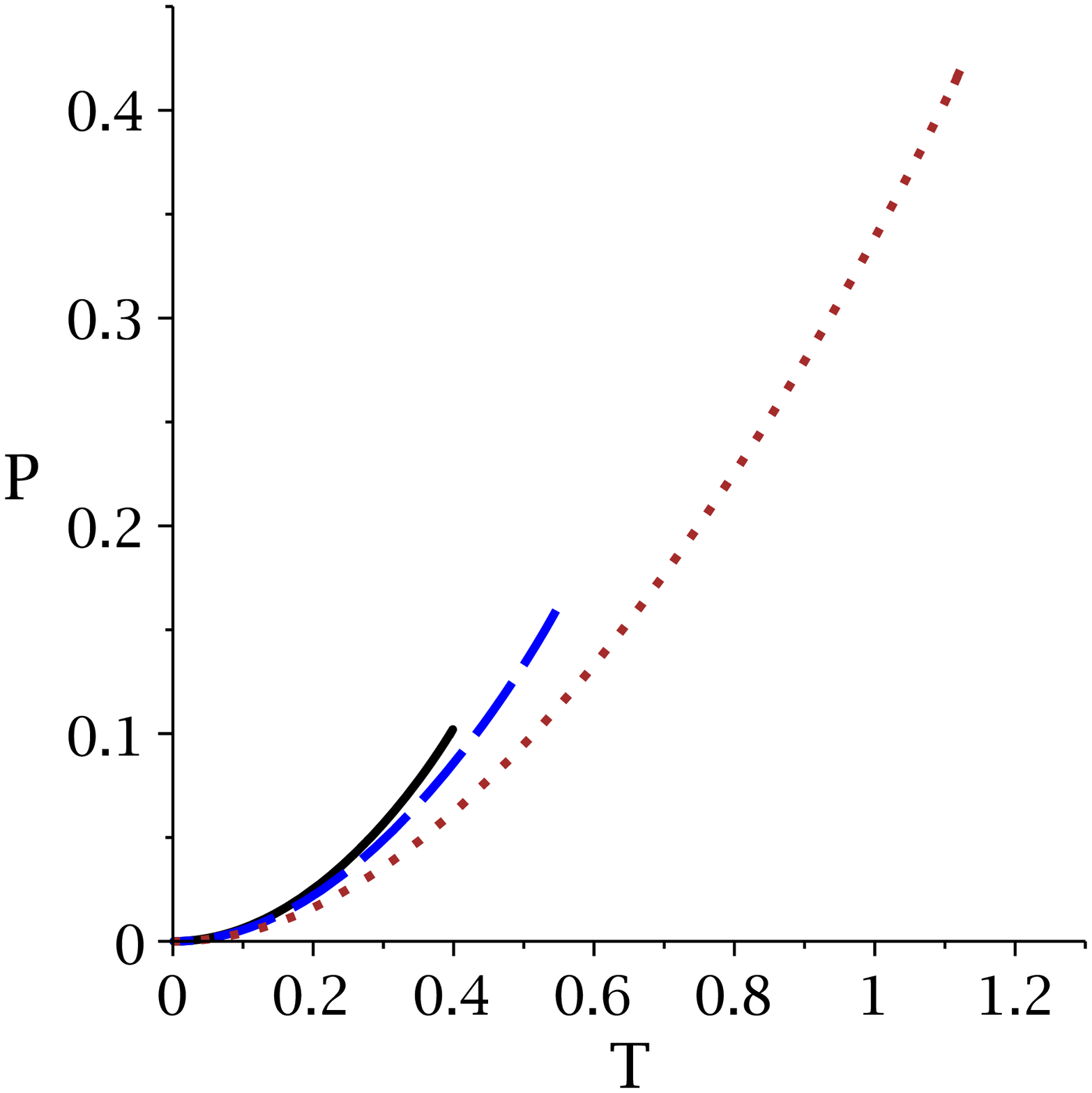}%
\end{array}
$%
\caption{$P$ versus $r_{+}$ (left panel) and $P$ versus $T$ (right panel)
for $c_{0}=c_{1}=c_{3}=c_{4}=2$, $c_{2}=3$ $k=1$, $q=1$ and $n=4$; $m=0$
(continues line), $m=0.1$ (dashed line) and $m=0.2$ (dotted line). }
\label{Fig11}
\end{figure}

\begin{figure}[tbp]
$%
\begin{array}{cc}
\epsfxsize=6cm \epsffile{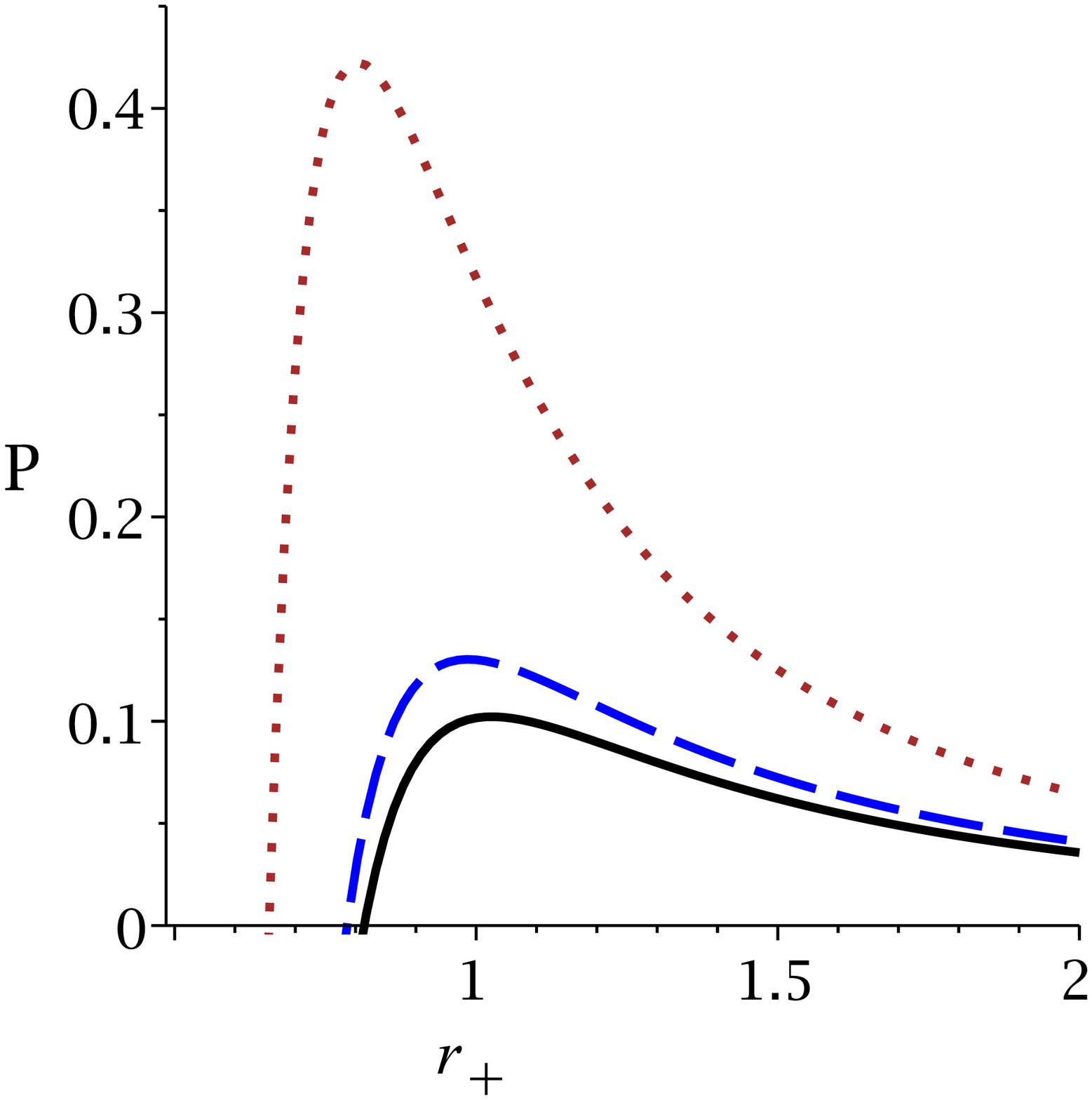} & \epsfxsize=6cm %
\epsffile{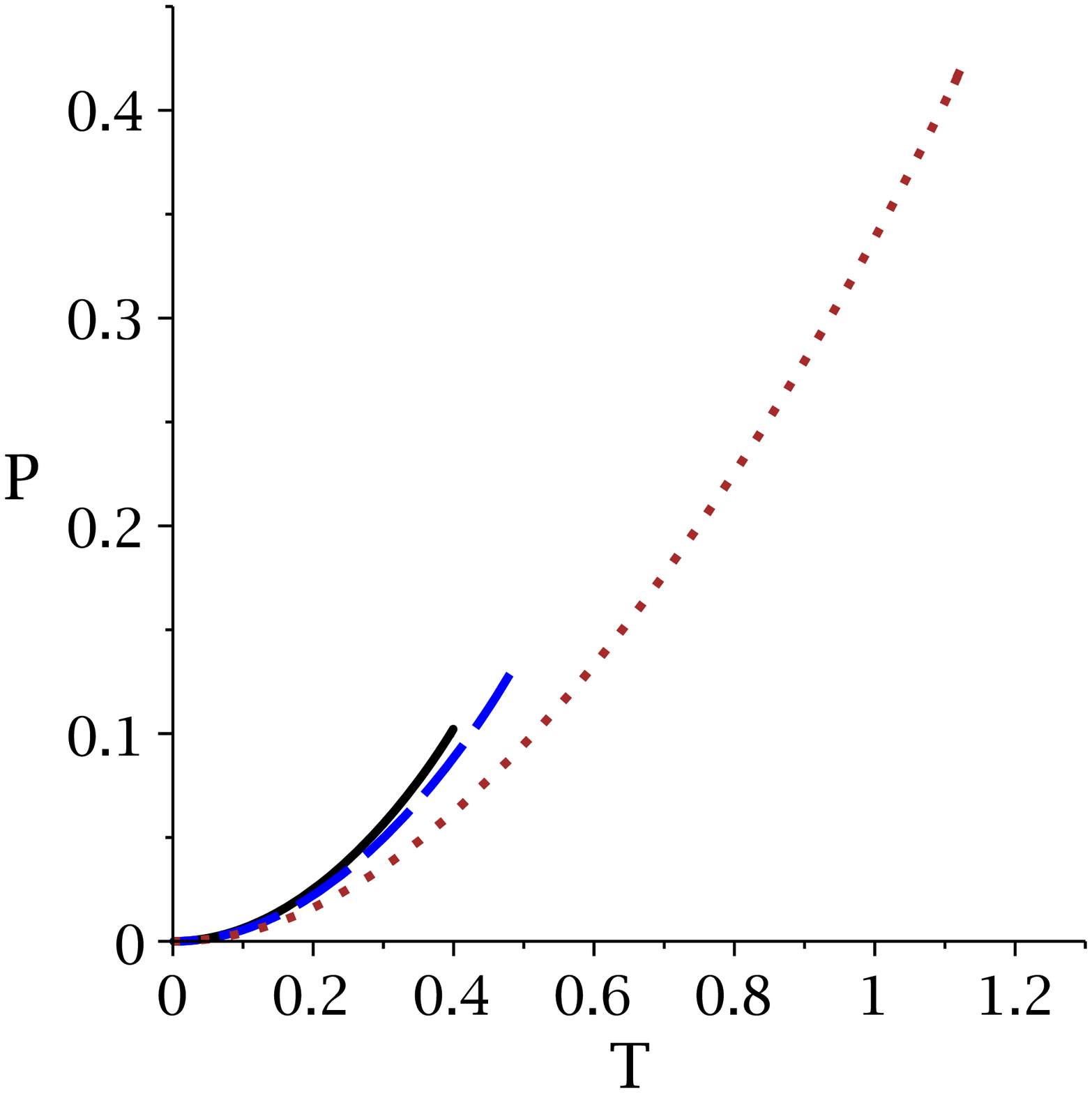}%
\end{array}
$%
\caption{$P$ versus $r_{+}$ (left panel) and $P$ versus $T$ (right panel)
for $c_{1}=c_{3}=c_{4}=2$, $c_{2}=3$ $k=1$, $q=1$, $m=0.2$ and $n=4$; $%
c_{0}=0.1$ (continues line), $c_{0}=1$ (dashed line) and $c_{0}=2$ (dotted
line). }
\label{Fig22}
\end{figure}

\begin{figure}[tbp]
$%
\begin{array}{cc}
\epsfxsize=6cm \epsffile{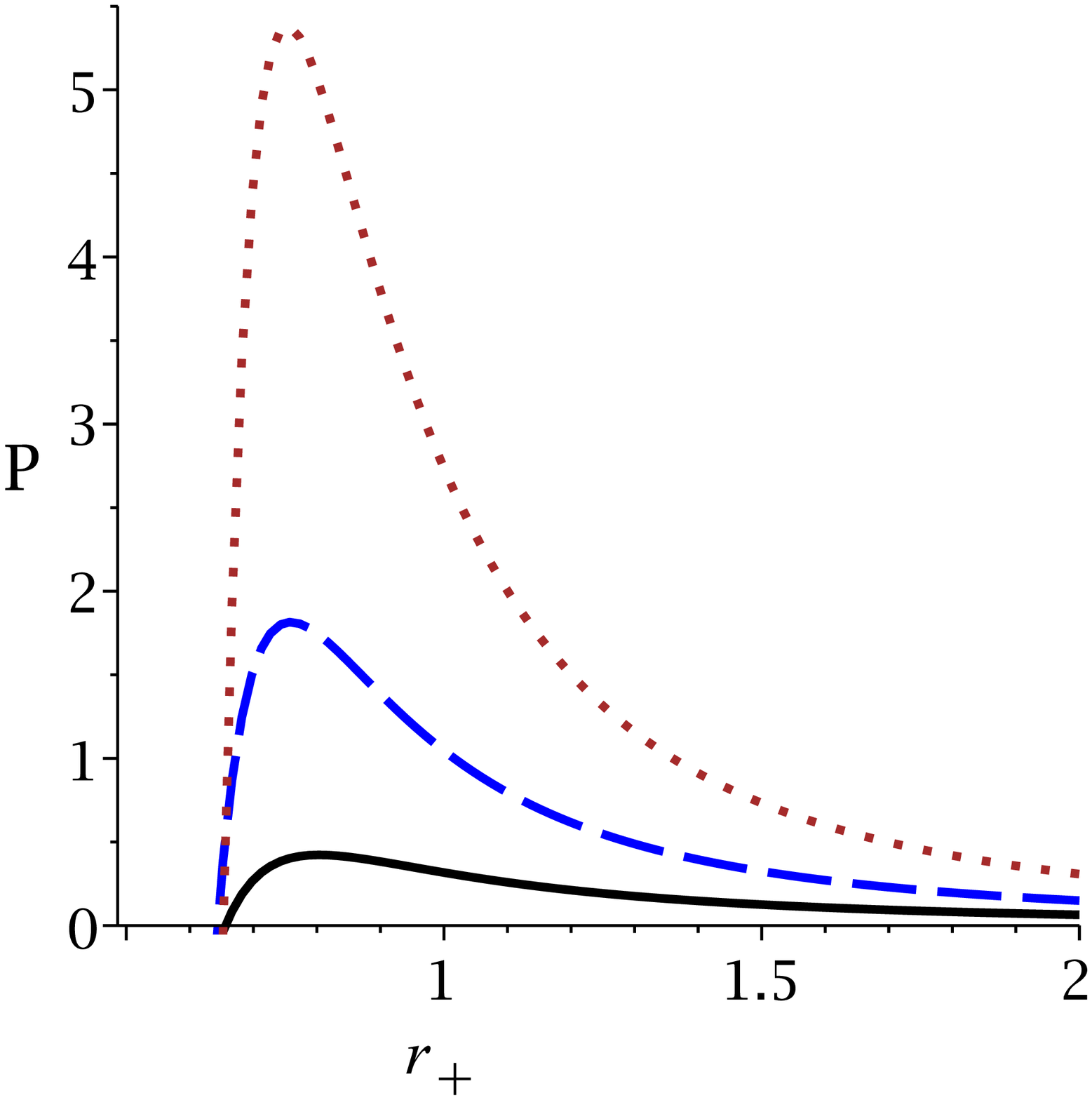} & \epsfxsize=6cm %
\epsffile{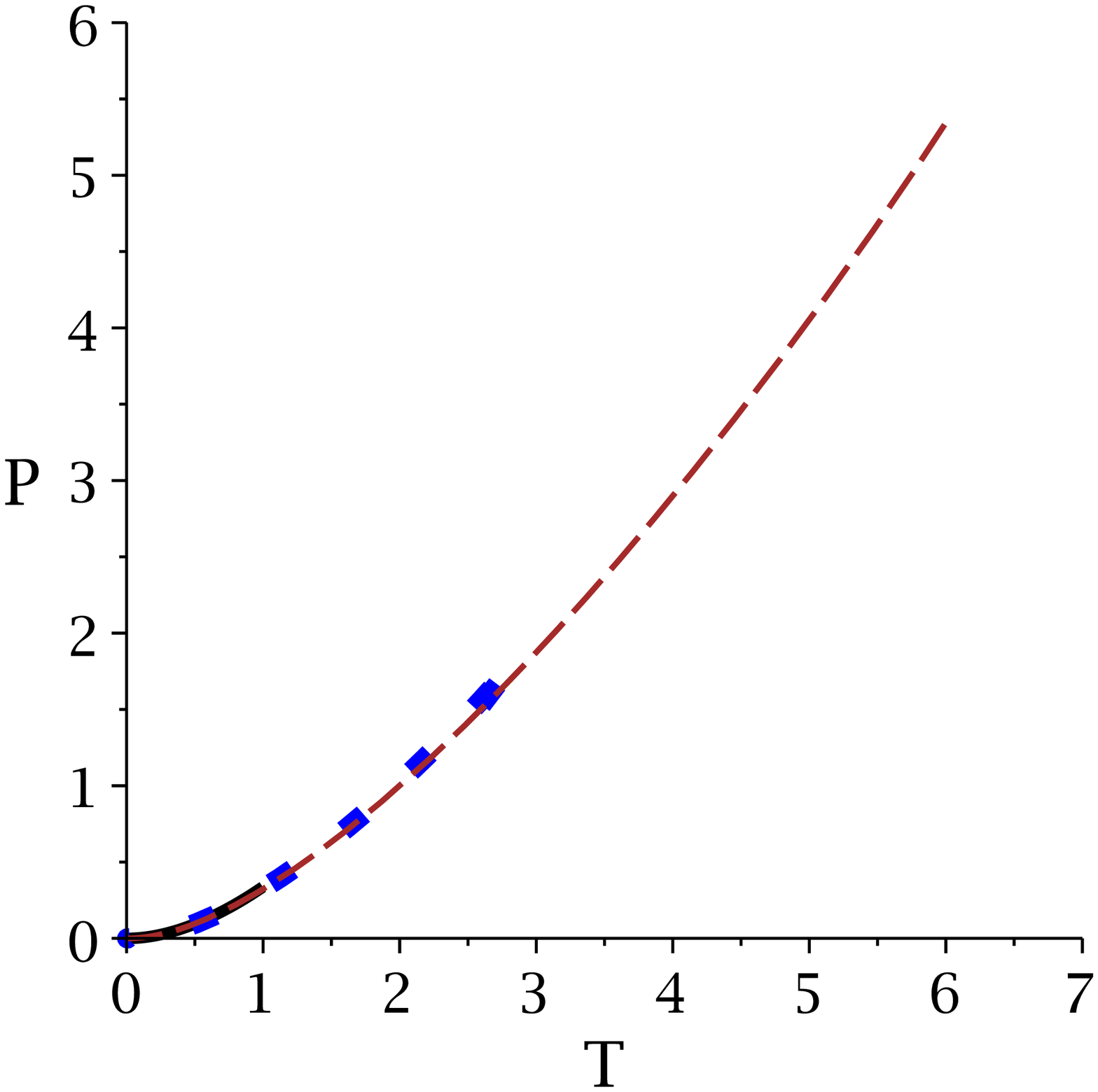}%
\end{array}
$%
\caption{$P$ versus $r_{+}$ (left panel) and $P$ versus $T$ (right panel)
for $c_{0}=c_{1}=c_{3}=c_{4}=2$, $c_{2}=3$ $k=1$, $q=1$ and $m=0.2$; $n=4$
(continues line), $n=5$ (dashed line) and $n=6$ (dotted line) in $P-r_{+}$
diagrams. \newline
$n=4$ (continues line), $n=5$ (dotted line) and $n=6$ (dashed line) in $P-T$
diagrams. }
\label{Fig33}
\end{figure}

In order to find the critical horizon radius, one should calculate first
order derivation of the pressure with respect to horizon radius and solve it
for $r_{+}$. Calculations show that it is not possible to obtain critical
horizon radius analytically. Therefore, we employ numerical method. The
results are presented in various diagrams (Figs. \ref{Fig11}-\ref{Fig33}).
In addition, we have plotted coexistence curve which characterizes two
different phases of small/larger black holes with same temperature and
pressure.

Evidently, due to the existence of maximum for pressure, these black holes
enjoy second order phase transition in their phase space. The critical
pressure and temperature are increasing functions of the massive parameter
(Fig. \ref{Fig11}), $c_{0}$ (Fig. \ref{Fig22}) and dimensions (Fig. \ref%
{Fig33}). As for critical horizon radius, it is a decreasing function of the
massive parameter (left panel of Fig. \ref{Fig11}) and $c_{0}$ (left panel
of Fig. \ref{Fig22}) while it is an increasing function of dimensions (left
panel of Fig. \ref{Fig33}). In addition, the length of coexistence curve is
an increasing function of the massive gravity, $c_{0}$ and dimensions.

The massive parameter, $m$, is a measurement for the mass of the
graviton. Here, we see that by increasing this parameter, the
critical pressure and temperature increase while critical horizon
radius decreases. This indicates that for massive graviton, black
holes have phase transition in smaller horizon radius. On the
other hand, considering massive graviton leads to increment of
length of coexistence line between two different phases of smaller
and larger black holes. Therefore, system in this case goes under
second order phase transition in higher pressure and temperature.
This effect highlights one of the differences between
gravitational systems with massive gravitons and those with
massless ones.

\section{Closing Remarks}

In this paper, we have reported a brief discussion regarding to
the black hole solutions of massive gravity with a linear $U(1)$
gauge field. One of the interesting results of this paper was
contribution of the massive gravity to number, type and place of
horizons. It was shown that the mentioned three properties are
highly sensitive to variation of the massive parameter and
geometrical mass. It was pointed out that considering different
values for the parameters will lead to different structures for
the black holes. Although the massive terms could be up to $n+2$
number, it is hard to study the total behavior of the solutions
with this number of terms. Hence, for the sake of simplicity, we
restricted our study to the fourth term.

From thermodynamical point of view, we have studied thermal
stability, phase transition and geometrical thermodynamic behavior
of the charged black holes in massive gravity. It was shown, in
case of finding a critical massive parameter for both root and
divergence points of the heat capacity, one will come across two
sets of equations. In case of the root, the presence of all
massive coefficients was only observed in the denominator of the
obtained relation which showed that root of the heat capacity
depends on these coefficients. Whereas for the critical massive
parameter and divergence points of the heat capacity, one of the
massive coefficients ($c_{1}$) had no contribution.

As for the physical/nonphysical point and second order phase
transitions, it was pointed out there are three cases; i) only one
bound point, ii) one bound point and one second order phase
transition, and finally iii) one bound point and two second order
phase transitions. These three situations are due to the
contributions of massive part of solutions. In other words,
considering the value of massive parameter, one of these cases may
happen. Interestingly, smaller divergence point of the heat
capacity for one dimension coincided with smaller divergence point
of the other dimensions. The place of coincidences were different
for every dimension. In other words, coincidence between two
different dimensions took place for the specific value of massive
parameter. For example, for $n=4$ there are different values of
the massive parameter (say $m_{1}$, $m_{2}$ and etc.) in which the
horizon radius for smaller second order phase transition will be
the same for $n=5$, $n=6$ and etc. Moreover, upon cancelling the
electric charge of the solutions this characterized behavior was
modified and the presence of second phase transition was vanished.
This behavior highlights the role of the electric charge in
thermodynamical behavior of the system.

Next, it was shown that thermal stability of the solutions is a
function of bound and second order phase transition points. In
other words, considering the number of the phase transition
points, system may have different stable/unstable states.

At last, we employed geometrical approach for studying the thermodynamical
behavior of the system. It was shown that employed thermodynamical metric in
this paper, provided a suitable machinery for studying the phase transition
of these black holes. In other words, regardless of bound point or second
order phase transition, all the divergence points of Ricci scalar of the
considered metric coincided with all root and divergence points of the heat
capacity. It was also seen that characteristic behavior of the Ricci scalar
enables one to recognize the bound point from the second order phase
transitions.

In addition, using thermodynamical concept, a study regarding critical
behavior of the system was conducted. The effects of different parameters on
critical values and coexistence curve between two phases were investigated.
The results regarding the massive parameter were investigated and
considerable effects of $m$ on the phase transition status were reported.

The main goal of this paper was investigation of geometrical
thermodynamics and van der Waals like phase transitions in the
context of a class of massive gravity. In order to obtain a
general model of massive gravity, one may follow Stueckelberg
mechanism and resuscitate the Goldstone modes which additionally
come from breaking diffeomorphism. For example, in Ref.
\cite{Poincare}, it was shown that relaxation of the Poincare
invariance leads to introduction of the large number of the
options for generical massive gravity models. It is also
worthwhile to consider a dilaton field in addition to Maxwell one
and study its consequences. Moreover one can apply various
nonlinear electrodynamics instead of Maxwell field in the context
of massive gravity. Generalization of Einstein theory to modified
and higher derivative gravities with a massive field is another
interesting subject. In addition, since obtained black hole
solutions have various number and type of horizons, it will be
interesting to study thermodynamical relations related to these
horizons \cite{Xu} and the possibility of the anti-evaporation
property \cite{anti1,anti2}. In addition, following the approach
of Ref. \cite{cosmology3}, it will be interesting to discuss the
effective couplings of action (\ref{Action}) with scalars to
obtain a consistent theory like bigravity. All these works could
be addressed elsewhere.

\begin{acknowledgements}
We are grateful to the anonymous referees for the insightful
comments and suggestions, which have allowed us to improve this
paper significantly. In addition, we would like to thank Matteo
Baggioli for useful discussions. We also thank Shiraz University
Research Council. This work has been supported financially by the
Research Institute for Astronomy and Astrophysics of Maragha,
Iran.
\end{acknowledgements}

\end{document}